\documentclass[twocolumn,aps,prb,raggedbottom,nobalancelastpage,amssymb,superscriptaddress]{revtex4-1}
\usepackage{amsmath}
\usepackage{amssymb}
\usepackage{amsfonts}
\usepackage{dsfont}
\usepackage{graphicx}
\usepackage{bm}
\usepackage{color}
\usepackage{appendix}
\usepackage{epsfig}
\usepackage{amsmath}
\usepackage{mathrsfs}
\usepackage{amssymb}
\usepackage{amsfonts}
\usepackage{dsfont}
\usepackage{graphicx}
\usepackage{bm}
\usepackage{color}
\usepackage{appendix}
\usepackage{epsfig}
\usepackage{tikz}
\usepackage{tikz}
\usetikzlibrary{shapes.geometric}
\usetikzlibrary{decorations.markings}
\usetikzlibrary{decorations.pathmorphing}
\usepgflibrary{decorations.shapes}
\usepgflibrary{shapes.misc}
\newcommand{\ZZ}{\mathbb{Z}}
\DeclareMathOperator{\Tr}{Tr}

\begin{document}
\title{Classical topological paramagnetism}

\author{R.~Bondesan}
\affiliation{Theoretical Physics, Oxford University, 1, Keble Road, Oxford OX1 3NP, United Kingdom.}
\author{Z.~Ringel}
\affiliation{Theoretical Physics, Oxford University, 1, Keble Road, Oxford OX1 3NP, United Kingdom.}

%%%%%%%%%%%%%%%%%%%%%%%%%%%%%%%%%%%%%%%%%%%%%%%%%%%%%%%%%%%%%%%%%%

\begin{abstract}
  Topological phases of matter are one of the hallmarks of quantum
  condensed matter physics. One of their striking features is a
  bulk-boundary correspondence wherein the topological nature of the
  bulk manifests itself on boundaries via exotic massless phases. In
  classical wave phenomena analogous effects may arise; however, these cannot be viewed as equilibrium phases of
  matter. Here we identify a set of rules under which robust equilibrium classical topological phenomena exist. We write down simple and analytically tractable classical lattice models of spins and rotors in two and three dimensions which, at suitable parameter ranges, are paramagnetic in the bulk but nonetheless exhibit some unusual long-range or critical order on their boundaries. We point out the role of simplicial cohomology as a means of classifying, writing-down, and analyzing such models. This opens a new experimental route for studying strongly interacting topological phases of spins.  \end{abstract}

%\pacs{73.20.-r, 74.25.Ha,  }

\maketitle
%\tableofcontents

\section{Introduction}

Symmetry protected topological phases are exotic quantum states of matter that are featureless in the bulk but still support unusual low energy phenomena on their boundaries. Their distinguishing properties remain sharp and robust as long as the appropriate symmetries are maintained.  An important example is the quantum spin Hall insulator \cite{Hasan2010}, protected by time reversal symmetry, whose edge physics may be used in spintronics~\cite{Ilan2014,Wu2011,Ojeda2012} and in the creation of topologically protected qubits in the form of Majorana fermions \cite{FuKane2008}. Partially motivated by the search for other exotic boundary phenomena, the field has developed rapidly: The classification table of weakly interacting topological phases of electrons given various symmetries has been established \cite{TenFold2010} in what can be seen as a modern revival of band structure theory. Furthermore, various topological electronic phases have been realized \cite{Hasan2010,Qi2011}. Turning to bosons, a difficulty arises since without interactions their ground state is always a superfluid regardless of the band structure. Nonetheless such phases do exist at strong interactions and are known as bosonic SPTs \cite{Chen2011,Chen2011a,Schuch2011}. Unfortunately, experimental realizations of bosonic SPTs are scarce and, to the best of our knowledge, limited to one dimensional spin chains \cite{Buyers1986}. 

Recently there has been both theoretical \cite{Gennady2013,Kane2014,
PhysRevLett.116.135503}
and experimental
\cite{Huber2015,PhotonicsExp1,PhotonicsExp2,PhotonicsExp3,Paulose23062015,Chen09092014} interest in
the notion of classical topological phases mimicking the
  phenomenology of their quantum counterparts. A typical strategy there is
to consider systems of springs and masses or optical devices which have an underlying topological band
structure.
% . Instead of populating these bands with electrons, one
% populates them with classical phonons or photons. 
Their edges can be seen as robust waveguides which have potential
engineering applications, such as delay lines for light and sound
\cite{Hafezi2011}. Notwithstanding, it is difficult to view such
phenomena as a distinct phase of matter, since the topological nature
of the band structure does not induce any sharp measurable features in
equilibrium. Further, at present the effect of non-linearities on
these systems is unclear. (See however \cite{Chen09092014}.)  Both
these issues can be seen as a classical reflection of the
aforementioned difficulty of finding topological equilibrium phases of
non-interacting bosons. As in the quantum case, an alternative route
may thus be to consider strongly interacting systems. 

 One approach to obtain such models is to start from known quantum SPT models and
attempt to write their discretized Euclidean time partition function
in a sign-problem free and local manner.  When possible, the resulting partition function can then be viewed as a classical statistical
mechanical system. Nonetheless, the models thus obtained have several
drawbacks.  First, the notion of symmetry protection does not
generally carry through to the classical problem, in the following
sense:  We define classical symmetries as those one-to-one maps on
configuration space which leave the Boltzmann weight invariant. For instance, in a spin-1
antiferromagnetic chain which supports a 1D SPT known as the Haldane
phase \cite{AKLT1988}, the associated classical configuration space is
one discrete variable ($m_z=-1,0,1$) per site. When viewed as an SPT
phase protected by $SO(3)$ or its $Z_2 \times Z_2$ subgroup of
$\pi$-rotations \cite{Pollmann2012}, the action of the symmetry involves superpositions and
cannot be considered classical.  A related issue is that the
microscopic mechanism stabilizing topological phases, based on matrix
product states and projective symmetries \cite{Schuch2011}, becomes
obfuscated in the classical setting. Lastly, the Boltzmann weights
resulting from the prescription outlined above, are complicated and
anisotropic, making these models less experimentally relevant.

Interestingly, for some models based on coupled superfluids, the lattice Euclidean time partition function, following a series of transformations, can be written in a sign free manner \cite{Geraedts2013}. The advantage here is that the resulting models are isotropic. However in the process of making the action local, additional degrees of freedom are introduced and, from a classical perspective, it is thus unclear what are the essential ingredients which render this a well defined classical phase of matter rather than a particular model. Furthermore it will be useful to generalize this approach to the discrete symmetry case which is more experimentally relevant.  
% Here we address the question of what restrictions should be imposed on a classical system, in any dimension, under which it supports robust classical topological phases (CTPs). We propose that the systems to consider are those which obey both a classical symmetry ($G$) and a local constraint whose defects can be labelled by a group $G'$. Furthermore the admissible phases are those which do not break the symmetry spontaneously and confine defects ($g'_1,g'_2 \in G'$) into neutral ($g'_1g'_2 = I$) pairs. Under these rules we find that there are many distinct topological phases, both in $2D$ and $3D$, with the accompanying exotic boundary phenomena. The latter include a ``forbidden" 
% \cite{VanHove1950} symmetry breaking order along $1D$  boundary and an unusual $2D$ critical phase
% corresponding to a theory of a compact boson in which the basic
% $\pm 2\pi$ vortices are linearly confined.

In this work we address the question of what restrictions, analogous
to symmetry protection, should be imposed on a classical system under
which it supports robust classical topological phases (CTPs). The first requirement is to
consider systems invariant under a group $G$ and a local constraint
whose defects carry elements of another group $G'$.  (More details about defects can
be found in Appendix \ref{App:Constraint}.) One example would be a
gauge theory with gauge group $G'$ and defects being monopoles. The
second requirement is that these phases must be short range correlated
in the bulk and in particular must not break the symmetry
spontaneously. The third is that they must confine defects of the
constraints into neutral pairs (see Appendix
\ref{App:Constraint} for a precise definition). We refer to phases
which obey the above restrictions as ``admissible phases''.

Interestingly, we find that given a dimension $d$, and the groups $G$,
and $G'$ as above, there are many inequivalent admissible phases. As
standard, two phases are deemed equivalent if a continuous deformation
from one to another is possible without crossing a critical point. By
continuous we mean that one deforms the energy functional gradually
and maintains the local constraint. We establish the existence of
inequivalent phases by providing concrete examples of models between
which any continuous deformation must involve a phase
transition. Notably, given that such distinct phases exist, by
definition their distinction does not involve a local order parameter
or confinement-deconfinement transitions. Their difference is of
topological origin. This is manifested on interfaces between distinct
phases, where either long range correlated or quasi long range
correlated phases emerge.

In the next sections we will explore these ideas for the choice 
$G=G'=Z_N$, considering models in both $2D$ and $3D$ where we find many distinct topological phases with the accompanying exotic boundary phenomena. The latter include a ``forbidden" \cite{VanHove1950} symmetry breaking order along $1D$ boundary and an unusual $2D$ critical phase corresponding
to a theory of a compact boson in which the basic $\pm 2\pi$ vortices
are linearly confined. Just as group cohomology was shown to be the basis for quantum SPTs
phases, we will show that tools from cellular cohomology \cite{Cohomology1991} give a powerful
mathematical framework for writing down models of CTPs and analyzing
them. The models thus produced are compact, isotropic and, to a large
extent, analytically tractable, thereby increasing both their theoretical
and experimental relevance.
%. 
The $G=G'=\ZZ_2$ models in $2D$ and $3D$ are further shown to be in
the same universality class as the imaginary time partition function
of certain $1D$ and $2D$ models (the group cohomology models
\cite{Chen2011}) of bosonic SPTs. From a numerical perspective our models thus provide an efficient way for performing Monte-Carlo simulations of bosonic SPTs with discrete symmetries (see also Ref.~\onlinecite{Geraedts2013} for the continuous case). They also open a new and more promising experimental route for studying these fascinating strongly interacting phases of matter.

\section{Two dimensions}

As a first illustrative example of a $2D$ CTP with $G=\ZZ_2$ we consider the
following model on the square lattice:
\begin{align}
   \label{eq:Z_sigma_A}
  Z &= \sum_{\sigma,U}e^{-\beta\mathscr{H} }
  \prod_{p}\delta(U_{ij}U_{jk}U_{kl}U_{li} - 1)\, ,\\
  \label{eq:model}
  -\beta\mathscr{H} &= 
  \sum_{\langle i,j\rangle}
  \left\{
    K_1 \sigma_i U_{ij}\sigma_j + K_2 U_{ij}\right\}\, .
\end{align}
Here $\sigma_i=\pm 1$ and $U_{ij}=\pm 1$ are site and link variables,
and the product is over plaquettes $p$, having the sites $i,j,k,l$ on
their boundary. The model has a $\ZZ_2$ symmetry
$\sigma_i\to -\sigma_i$, and it has a $\ZZ_2$ constraint forcing zero
flux for the $U$ field through each plaquette.

Conveniently, a non-local transformation ($U_{ij}=\mu_i\mu_j$) maps this model to two decoupled Ising models,
and has thus $\ZZ_2\times \ZZ_2$ symmetry:
\begin{align}
  \label{eq:Zsmu}
  Z
  &=
  \sum_{\sigma,\mu}
  \exp
  \sum_{\langle i,j\rangle}
  \left\{
    K_1 \rho_i\rho_j + 
    K_2 \mu_i\mu_j\right\}\, ,
  &
  \rho_i &= \sigma_i \mu_i
  \, .
\end{align}
Denoting $K_c = -\frac{1}{2}\log\tanh K_c$ the critical coupling of
the Ising model on a square lattice, there are two regimes which are
of interest to us: The trivial phase ($K_2>K_c>K_1\ge 0$) where
$\langle \mu_i\rangle\neq 0$, and the non-trivial phase
($K_1>K_c>K_2\ge 0$) where $\langle \rho_i\rangle\neq 0$.  The other
variables, $\rho$ and $\sigma$ for the trivial case and $\mu$ and
$\sigma$ for the non-trivial, are disordered. Notably, in both cases
$U_{ij}$'s are uncorrelated, namely
$\left\langle (U_{ij}-\langle U_{ij} \rangle) (U_{kl}-\langle U_{kl}
  \rangle) \right\rangle$ is exponentially decaying.  \footnote{We
  remark that one can consider more general couplings, such as those
  of the Ashkin--Teller model \cite{Baxter}, as long as the set of
  order parameters in a phase is unchanged. Our choice of two
  decoupled Ising models is made for pedagogical purposes.}  We remark
that the partition function of Eq. (\ref{eq:model}) with constraint
violations at two plaquettes equals that of Eq.~\eqref{eq:Zsmu} where
the sign of both couplings $K_1,K_2$ is reversed along a path
connecting the two plaquettes \cite{Savit}. Thus for both regimes, the
presence of order parameters with long range order implies linear
confinement of the defects.

In terms of $\rho$ and $\mu$, the model is simply two decouple
ferromagnets that exhibit symmetry broken phases.  However, in the
original degrees of freedom, $U,\sigma$, the physical properties of
the two phases change. Considering bulk physics, long range order in
$\rho$ implies the following non local (string) order parameter in the
non-trivial phase:
\begin{align}
  \label{eq:sAs}
   \left< \rho_i \rho_j \right> &=\left< \sigma_i\prod_{\ell \in \Gamma_{ij}}U_{\ell} \sigma_j\right> 
  \to \text{const}\, 
\end{align}
as $\text{dist}(i,j)\to\infty$ and $\Gamma_{ij}$ is a path from $i$ to
$j$.  Alternatively stated, performing the non-local transformation
$\sigma_i \rightarrow \rho_i=\prod_{\ell\in\Gamma_{0i}}U_\ell
\sigma_i$, with $0$ a reference site, unveils a hidden ferromagnetic
phase for the non-trivial order, whereas for the trivial phase, this
results in a simple paramagnet. 

As we now argue the hidden ferromagnetic order is a distinguishing property of the topological phase and therefore one cannot continuously deform the models onto one another. This implies that there are at least two distinct admissible phases in our classification for $d=2,G=G'=Z_2$. Notably local and symmetric
perturbations in the original $U$ and $\sigma$ variables would be
transformed into local and symmetric perturbations in $\mu$ and
$\sigma$. As this transformation has no effect on the free energy, one
finds that hidden order is thermodynamically equivalent to
conventional order. This means that hidden order not just a feature of the model but rather a robust property which can only vanish through a phase transition or by leaving the space of admissible phases. 

Perhaps the most interesting distinction between these two phases
comes about when considering a $1D$ interface between them. In
general, near an interface between a ferromagnet and a paramagnet, the
order parameter leaks into the paramagnetic phase up to some penetration
length. Similarly, close to an interface between the above two phases
both order parameters ($\rho$ and $\mu$) will be ordered and as
a result $\sigma = \rho\cdot \mu$ would also be ordered,
despite being disordered in the bulk on both sides. For instance,
setting $K_1=0,K_2\to \infty$ on the trivial side is equivalent to
placing the non-trivial phase in an open geometry with boundary
conditions $U_{ij}=1$ or equivalently $\mu_i=\mu_j$, implying long
range order for $\sigma$. 

\begin{figure}[ht!]
   \newcommand{\Height}{2}
 \newcommand{\Width}{4}
 % \begin{figure}[ht!]
%   \newcommand{\Height}{2}
% \newcommand{\Width}{4}
% \input{2D_conf.tex}
%   \caption{Pictorial representation of low energy configurations of
%     the $2D$ classical topological paramagnet.  
%     Red lines are domain walls of the spins, while the blue ones are those
%     where the link variable $U=-1$. 
%     In the bulk both of these lines must form closed paths and 
%     energetically they are also encouraged to pair up, and
%     at a boundary (bottom) $\sigma$ has long range order.}
% \label{fig:conf}
% \end{figure}

\begin{tikzpicture}[scale=1,thick]

\coordinate (O) at (0,0);
\coordinate (A) at (\Width,0);
\coordinate (B) at (\Width,\Height);
\coordinate (C) at (0,\Height);

\draw[fill=red!20,opacity=0.8] (O) -- (A) -- (B) -- (C) -- cycle; % rect
%\draw[yellow!80] (O) -- (A); % Bottom line

%legend
\begin{scope}[yshift=.5cm]
\begin{scope}[xshift=5cm,yshift=0cm,scale=.5]
\node at (.5,2.5) {$U$};
\draw (0,0) rectangle (1,1);
\node[inner sep=0pt,minimum size=.1cm,circle,fill=black] at (0,0) {};  
\node[inner sep=0pt,minimum size=.1cm,circle,fill=black] at (1,0) {};  
\node[inner sep=0pt,minimum size=.1cm,circle,fill=black] at (1,1) {};  
\node[inner sep=0pt,minimum size=.1cm,circle,fill=black] at (0,1) {};  
\draw[blue] (.5,-.5)--(.5,.5)--(-.5,.5);
\node[label=below:{{\footnotesize $-1$}}] at (.5,0) {};
\node[label=left:{{\footnotesize $-1$}}] at (0,.5) {};
\node[label=right:{{\footnotesize $1$}}] at (1,.5) {};
\node[label=above:{{\footnotesize $1$}}] at (.5,1) {};
\end{scope}
\begin{scope}[xshift=6.5cm,yshift=0cm,scale=.5]
\node at (.5,2.5) {$\sigma$};
\draw (0,0) rectangle (1,1);
\node[inner sep=0pt,minimum size=.1cm,circle,fill=black,label=below:
{{\footnotesize$\uparrow$}}] at (0,0) {};  
\node[inner sep=0pt,minimum size=.1cm,circle,fill=black,label=below:
{{\footnotesize $\downarrow$}}] at (1,0) {};  
\node[inner sep=0pt,minimum size=.1cm,circle,fill=black,label=above:
{{\footnotesize $\uparrow$}}] at (1,1) {};  
\node[inner sep=0pt,minimum size=.1cm,circle,fill=black,label=above:
{{\footnotesize $\uparrow$}}] at (0,1) {};  
\draw[red] (.5,-.5)--(.5,.5)--(1.5,.5);
\end{scope}
\end{scope}

\def \a {3}
\def \b {2}
\def \c {4}
\pgfmathsetmacro{\n}{\a+\b+\c}
\foreach \s in {0,...,\a}
{
  \node[draw, circle, fill=black, inner sep=0pt,minimum size=.1cm,
  label=below:{$\uparrow$}] at (\Width/\n * \s,0) {};
}
\pgfmathsetmacro{\i}{\a+1}
\pgfmathsetmacro{\j}{\a+\b}
\foreach \s in {\i,...,\j}
{
  \node[draw, circle, fill=black, inner sep=0pt,minimum size=.1cm,
  label=below:{$\downarrow$}] at (\Width/\n * \s,0) {};
}
\pgfmathsetmacro{\i}{\a+\b+1}
\pgfmathsetmacro{\j}{\n}
\foreach \s in {\i,...,\j}
{
  \node[draw, circle, fill=black, inner sep=0pt,minimum size=.1cm,
  label=below:{$\uparrow$}] at (\Width/\n * \s,0) {};
}

%bulk
\begin{scope}[xshift=-1.5cm,yshift=-.2cm]
\draw[red!65!white] (3,1.5) ellipse (.75cm and .4cm);
\draw[blue!65!white] (3,1.5) ellipse (.85cm and .5cm);            
\end{scope}
%bdry
\begin{scope}[rotate=90,xshift=-.5cm,yshift=-3.125cm]
\draw[red!65!white] (.5,1.5) to[out=40, in=-40, looseness=3] (.5,.75);  
\end{scope}
\end{tikzpicture}
   \caption{Pictorial representation of low energy configurations of
     the $2D$ classical topological paramagnet.  
     Red lines are domain walls of the spins, while the blue ones are those
     where the link variable $U=-1$. 
     In the bulk both of these lines must form closed paths and 
     energetically they are also encouraged to pair up, and
     at a boundary (bottom) $\sigma$ has long range order.}
 \label{fig:2D_conf}
 \end{figure}
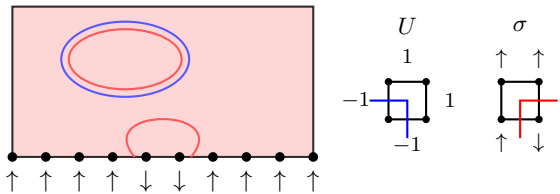

More physically, one can view the configurations of $U$ in
\eqref{eq:Z_sigma_A} as polygons on the dual lattice by assigning a
line of the polygon to links across which $U=-1$.  The $K_1$ coupling
then encourages domain walls of the spins to attach to these
polygons. Kinks of $\sigma$ along the interface are necessarily ends
of domain walls in the bulk. However these domain walls cannot have an
accompanying polygon as the latter is confined from entering the
trivial phase (vacuum in the picture). Consequently the bulk, despite
being locally disordered, linearly confines kinks of $\sigma$ at the
boundary into neutral pairs (see Fig. \ref{fig:2D_conf}).

\subsection{Relation with the AKLT Hamiltonian}
\label{sec:AKLT}
We now establish a precise connection between the $2D$ CTP presented
and the AKLT model, the paradigmatic example of a quantum SPT phase of
spins in $1+1D$ \cite{AKLT1988}.  (See also
\cite{chen2014symmetry} for a picture of AKLT that is close to our
construction.)  We consider the transfer matrix of the $2D$ CTP in the
limit of anisotropic coupling
$K^x_i=\epsilon \lambda_i\, , e^{-2K_i^y}=\epsilon \lambda_i'$,
$i=1,2$, along the horizontal ($x$) or vertical direction ($y$).  It
is then a standard exercise (see e.g.~\cite{Kogut}) to derive the
quantum Hamiltonian in the limit $\epsilon\to 0$ starting from Eq.~\eqref{eq:Zsmu}
in the main paper, and to pass from the $\mu$ variables to their duals
$\tau$. This results in the $\ZZ_2\times \ZZ_2$ symmetric Hamiltonian
$H=H_0+\sum \lambda_2\tau_{i+1/2}^x + \lambda_1'\sigma_i^x$, where
\begin{align}
  H_0
  =
  \sum
  \lambda_1
  \sigma_i^z\tau_{i+1/2}^x\sigma_{i+1}^z 
  +
  \lambda_2'\tau_{i-1/2}^z\sigma_i^x\tau_{i+1/2}^z \, ,
\end{align}
and which coincides with the AKLT Hamiltonian in the form considered
in \cite{Ringel2015} for $\lambda_1=\lambda_2'$. Having equivalent
phenomenology and a very similar algebraic structure strongly suggests
that these two models describe the same phase. Interestingly, when
expressing our model in terms of the dual variables $\tau$, the
Boltzmann weights are not positive anymore.  The $\ZZ_2$ constraint
thus appears as a natural way to reflect the additional $\ZZ_2$
symmetry while maintaining positive Boltzmann weights and locality.

\subsection{Generalizations to $G=G'=Z_N$}
\label{sec:gen_2d}

Let us generalize the above model to the case $G=G'=Z_N$.
Accordingly, we consider a directed square lattice and take
$\sigma_i \in {\rm Z}_N$ and $U_{ij} = U_{ji}^{-1} \in {\rm Z}_N$ for
the orientation being from vertex $i$ to $j$.  We represent elements in
${\rm Z}_N$ by $e^{2\pi i \alpha/N}$, $\alpha= 0,1,\dots,N-1$.  
For a given $p=0,1,\dots,N-1$, let us define the minimal coupling:
\begin{align}
  \mathscr{H}_{p} 
  &= 
  \sum_{i}\sum_{j\sim i}
  \sigma^p_i U_{ij} \sigma^{-p}_j\, ,
\end{align}
where $j\sim i$ means $j$ a neighbour of $i$, so that each edge is
counted twice, once with its positive and once with its negative
orientation ensuring a real energy. Given non-zero $p\neq p'$ the generalized model is
defined by \eqref{eq:Z_sigma_A} with
\begin{align}
  \label{eq:model_ZN}
  -\beta\mathscr{H}_{p,p'} &= 
  K_1  \mathscr{H}_{p} 
  +
  K_2 \mathscr{H}_{p'} 
  \, .
\end{align}
As we will show, for large $K_1$ ($K_2$) p ($p'$) controls the topological index.  Let us note that $\sigma^p$ is a
$\ZZ_N$ variable only when $p$ and $N$ are co-prime. Otherwise, it has
a reduced order, given by $N/p$.  In order to keep the physical
message of this section clear and concise, we do not delve here in these
number theoretic considerations, and assume $N$ to be prime.

To analyze the model we first expose the hidden order. To this end we
resolve the constraint using
\begin{align}
U_{ij} &= \mu_i \mu^{-1}_j
\end{align}
yielding
\begin{align}
  \mathscr{H}_{p} 
  &= 
  \sum_{i}\sum_{j\sim i}
  \sigma^p_i \mu_i \mu^{-1}_j \sigma^{-p}_j\, ,
\end{align}
and 
\begin{align}
  Z &= \frac{1}{N}\sum_{\sigma,\mu}e^{-\beta\mathscr{H}}\, ,
\end{align}
where the factor of $\frac{1}{N}$ comes from the $1$ to $N$ mapping
between $U_{ij}$ which respect the constraint and $\mu_i$.

Next we wish to go to the composite variables 
\begin{align}
\tilde{\sigma}_{i;p} = \mu_i \sigma^p_i\, ,\quad
\tilde{\mu}_{i;p'} = \mu_i \sigma_i^{p'} \, .
\end{align}
The assumption of $N$ prime guarantees that they are in $\ZZ_N$, and 
the assumption of $p\neq p'$ and a non-zero $p$ guarantees the mapping to be
invertible. The indices $p,p'$ make explicit the dependence on $p$ 
and $p'$ in the definition of $\tilde{\sigma}$ and $\tilde{\mu}$.

% where $(p)^{-1}$ is defined as the smallest positive number such that $[\exp(2\pi i /N)]^{p' (p')^{-1}} = \exp(2\pi i /N)$ which implies for instance $[\sigma_i]^{p p' (p)^{-1}}=\sigma^{p'}_i$. The inverse transformation from $(\tilde{\sigma},\tilde{\mu})$ to $(\sigma,\mu)$, when it exists, can be determined from  
% \begin{align}
% \sigma^{p-p'}_i &= \tilde{\sigma_i} \tilde{\mu_i}^{-1} \\ \nonumber 
% \mu^{1-p(p')^{-1}}_i &= \tilde{\sigma_i} \tilde{\mu_i}^{-p(p')^{-1}}
% \end{align}
% and for the $p'=0$ case we take 
% \begin{align}
% \tilde{\mu} &= \mu
% \end{align}

% For simplicity let us from now on focus on prime $N$. Given this the necessary and sufficient conditions for an inverse to exists are $p-p' \,\ mod \,\ N \neq 0$. Notably $p (p')^{-1} = 1$ implies $p=p'$. 

We thus find two decoupled $Z_N$ clock models,
\begin{align}
  -\beta\mathscr{H}_{p,p'} &= 
  \sum_i\sum_{j\sim i}
  \left(K_1 \tilde{\sigma}_{i;p}  \tilde{\sigma}_{j;p}^{-1} 
  + 
  K_2 \tilde{\mu}_{i;p'}\tilde{\mu}_{j;p'}^{-1}\right)\, ,
\end{align}
one in the composite variable $\tilde{\sigma}_p$ and the other in the
composite variable $\tilde{\mu}_{p'}$.  Now we suppose that the couplings
are such that one of the two variables, say $\tilde{\sigma}_p$, is
ordered (recall that if $N$ is prime, $\ZZ_N$ models can have only a
single symmetry broken phase), and that $\mu$ is disordered.
Notably, since $\tilde{\mu}_{p'} = \mu \sigma^{p'}$ this also implies
that $\tilde{\mu}_{p'}$ is disordered for all $p' \neq p$.  We then claim that under these
conditions the model is in a ``topological phase of type $p$''. Three
questions need to be answered to justify this statement: (i) why is
this a phase (ii) why do different $p$'s correspond to distinct phases
and (iii) why are they topological.

Considering the first point note that the hidden order of the $\sigma$
variables manifested by order in $\tilde{\sigma}_p$ is a robust
property. Indeed as argued in the previous case of an Ising
symmetry, any local symmetric and defect-free perturbation in
original model would map to a local term in the $\tilde{\sigma}_p$
and $\tilde{\mu}_{p'}$ degrees of freedom. Thus robustness of the
topological phase is implied by the usual robustness of broken
symmetry states. Turning to the second point, and the role of $p$,
we can simply note that two different values of $p$ correspond to two 
different order parameters and thus two different phases. Indeed if $\tilde{\sigma}_{i;p}$ is long range ordered then $\tilde{\sigma}_{i;p'}$ must be disordered as it is equal to a power of $\tilde{\sigma}_{i;p}$ times a non-trivial power of the disordered variable $\mu_i$.  
%Indeed note that if $\tilde{\sigma}=\mu_i \sigma^p_i$ is ordered then
%$\mu_i \sigma^{p'}_i=\mu_i^{1-p'(p)^{-1}} \tilde{\sigma}^{p' (p)^{-1}}$. Under our assumption $\mu_i^{1-p'(p)^{-1}}$ is never
%$(\mu_i)^0$ hence $\mu_i \sigma^{p'}_i$, the hidden order parameter of
%the $p'$ topological phase, gets disordered by $\mu_i^{1-p'(p)^{-1}}$.

Lastly, we justify the nomenclature topological. By this we mean that an interface between two distinct admissible phases would contain some form of long range or quasi long range order. Consider such an interface between a $p$ topological phase and a $p'$ topological phase. This scenario can be engineered by setting $K_2=0$ and $\tilde{\sigma}_p$ ordered on one side of the interface, and $K_1=0$ and $\tilde{\mu}_{p'}$ ordered on the other. On the interface these two order parameters will leak and so $(\mu_i \sigma^p_i)(\mu_i \sigma^{p'}_i)^{-1} = \sigma^{p-p'}_i$ would be ordered. Notably the latter, and only the latter, is a local order parameter and thus we have shown the existence of 1D long range order on such interfaces 

\section{Three dimensions}

Next we wish to generalize the above construction to $3D$. In $2D$ we
attached closed polygons to domain walls of the spins. Turning to
$3D$, polygons on the dual lattice appear naturally in $\ZZ_2$ gauge
theories, where they correspond to discrete flux lines. However
domains walls become $2D$ objects, and we instead look for a property of
the spins that can also be described in terms of polygons.

Such a spin quantity has been studied recently in
\onlinecite{Ringel2015,Scaffidi2016} and can be thought of as an
algebraic generalization of the usual continuum notion of
vorticity. Consider a cubic lattice and orient links and plaquettes.
Next place a spin variable $\sigma=\pm 1$ at each vertex. The discrete
vorticity $\omega_p$ on a plaquette $p$ is defined as
\begin{align}
\label{eq:omegap}
\omega_p = 
\frac{1}{2}
\sum_{(ij)\in \partial p}\epsilon_{ij}^p\frac{1-\sigma_i\sigma_j}{2}
\, ,
\end{align}
where the sum is over links on the boundary of $p$ and
$\epsilon_{ij}^p=1$ if the link is oriented as the plaquette, and $-1$
otherwise. We remark that $\omega_p=0,\pm 1$ and the choice of
plaquette orientation has no effect on the $\mathbb{Z}_2$ quantity
$(-1)^{\omega_p}$ that we consider below. For definiteness we choose
orientations as in figure \ref{fig:conf_omega0}.

An intuitive view on discrete vorticity comes form thinking of the
spins $\sigma_i =+1,-1$ as the elements $0,1$ in $\mathbb{Z}_2$. Then
$\omega_p$ appears as the discrete integral (i.e.~a sum) around a
plaquette over the discrete derivatives
$\frac{1}{2}(1-\sigma_i\sigma_j) \in\mathbb{Z}_2$. Here it is important
to interpret the discrete derivative as a variable in $\mathbb{Z}$ rather
than in $\mathbb{Z}_2$, and hence this sum can be non-zero multiple of
$|\mathbb{Z}_2|=2$. This is analogous to what one does when calculating
vorticity of a U$(1)$ variable ($\phi$) where derivatives
($i \phi^{-1} \partial_l \phi$) are taken in U$(1)$ but then
integrated over as elements in $\mathbb{R}$ whose sum can now be a non-zero
multiple of $2\pi$. 

In analogy with usual vorticity, the discrete vorticity obeys a
discrete version of the zero divergence constraint: Given any box on
the square lattice, $\sum_{p\in\text{box}}\omega_p = 0\mod 2$. This
can be shown by noting that for each box we can choose a clockwise
orientation (when looking from inside the box) on each
plaquette. Consequently, each link on the box would appear exactly
twice with opposite values of $\epsilon_{ij}^p$. Therefore discrete
vorticity lines form polygons on the dual lattice which obey the exact
same branching rules as fluxes in a $\ZZ_2$ gauge theory. 

Tools from lattice gauge theory, specifically cellular and simplicial cohomology,
shed further light on this quantity.  A thorough discussion of these
aspects are relegated below in section \ref{sec:cohom} where they will
be used to define discrete vorticity for other abelian groups.
% In that language $\sigma$, the
% matter field (a $0$-chain) gets mapped by a discrete derivative (the
% coboundary operator) to $\frac{1}{2}(1-\sigma_i\sigma_j)$ which is a
% flat gauge field (a $1$-cocycle). The discrete vorticity (technically,
% a Bockstein homomorphism) maps this $1$-cocycle onto a flux
% configuration which is divergentless and there exists a gauge
% configuration with matching fluxes (a trivial $2$-cocycle). These
% relations, detailed in the Supp.~Mat.~II, imply that for each $\sigma$
% configuration the discrete vorticity lines form polygons on the dual
% lattice which appear as fluxes of some gauge field configuration.

Armed with the notion of discrete vorticity and its properties, we can now introduce the $3D$ model. Consider spins $\sigma_i$ on the vertices
of a cubic lattice and $\ZZ_2$ gauge variables $A_{ij}$ on the links,
and choose the following energy
\begin{align}
  \label{eq:E_3D}
  -\beta\mathscr{H}
  = J_1 \sum_{p} (AAAA)_p + J_2 \sum_{p} (-)^{\omega_p} (AAAA)_p\, ,
\end{align}
with $(AAAA)_p$ being the product of the four $A_{ij}$
surrounding the plaquette $p$. 

In analogy with our $2D$ analysis we would now want to perform some non-local transformation to decouple the gauge variables from the
spins. Even though both flux and vorticity lines form closed polygons,
the number of distinct flux configurations, which spans all such
polygons, is bigger than that of vorticity configurations which only
span a subset. Therefore, for any vorticity there exists a matching
flux although the converse is not true. It follows that there
exists $A_{\sigma}$ such that
$(A_{\sigma}A_{\sigma}A_{\sigma}A_{\sigma})_p =
(-)^{\omega_p}$. Defining $\tilde{A} = A A_{\sigma}$, we obtain
 \begin{align}
   \label{eq:Atilde}
  -\beta\mathscr{H}
 &= J_1 \sum_{p} (-)^{\omega_p} (\tilde{A}\tilde{A}\tilde{A}\tilde{A})_p + J_2 \sum_{p} (\tilde{A}\tilde{A}\tilde{A}\tilde{A})_p \, .
\end{align}

\begin{figure}[ht!]
\newcommand{\Depth}{4}
\newcommand{\Height}{1.5}
\newcommand{\Width}{1.5}
  \begin{tikzpicture}[scale=1,thick]

\coordinate (O) at (0,0,0);
\coordinate (A) at (0,\Width,0);
\coordinate (B) at (0,\Width,\Height);
\coordinate (C) at (0,0,\Height);
\coordinate (D) at (\Depth,0,0);
\coordinate (E) at (\Depth,\Width,0);
\coordinate (F) at (\Depth,\Width,\Height);
\coordinate (G) at (\Depth,0,\Height);

\draw[fill=yellow!80] (O) -- (C) -- (G) -- (D) -- cycle;% Bottom Face
\draw[fill=blue!30] (O) -- (A) -- (E) -- (D) -- cycle;% Back Face
\draw[fill=red!10] (O) -- (A) -- (B) -- (C) -- cycle;% Left Face
\draw[fill=red!20,opacity=0.8] (D) -- (E) -- (F) -- (G) -- cycle;% Right Face
\draw[fill=red!20,opacity=0.6] (C) -- (B) -- (F) -- (G) -- cycle;% Front Face
\draw[fill=red!20,opacity=0.8] (A) -- (B) -- (F) -- (E) -- cycle;% Top Face

%legend
\node at (\Depth-1.5,-.4,0) {{\footnotesize boundary}};
\draw[thin] (\Depth+.75, \Height-1)--(\Depth+.75, \Height-.5)--
 (\Depth+.85, \Height-.4)--(\Depth+.85, \Height-.9)--
 (\Depth+.75, \Height-1);
\node at (\Depth+.6, \Height-.5) {{\tiny $p$}};
\draw[red] (\Depth+.5, \Height-.75)--(\Depth+1, \Height-.75) node[right,black] 
{{\footnotesize $: (-)^{\omega_p}=-1$}};

\begin{scope}[yshift=-.75cm]
  \draw[thin] (\Depth+.75, \Height-1)--(\Depth+.75, \Height-.5)--
 (\Depth+.85, \Height-.4)--(\Depth+.85, \Height-.9)--
 (\Depth+.75, \Height-1);
\node at (\Depth+.6, \Height-.5) {{\tiny $p$}};
\draw[blue] (\Depth+.5, \Height-.75)--(\Depth+1, \Height-.75) node[right,black] 
{{\footnotesize $: (AAAA)_p=-1$}};

\end{scope}

%     %surface
%     \fill [fill=gray,path fading=fade right]
%     (0,0)--(0,2)--(1,2.5)--(1,.5)--cycle;
%     \draw[thin,path fading=east] (0,0)-- ++(0,1pt);
% %    \draw[thin,path fading=east] (0,2)--(6,2);
% %    \draw[thin,path fading=east] (0,2)--(6,2);
%     %bdry
\begin{scope}[xshift=-.25cm]
        \begin{scope}[rotate=90,scale=0.75,xshift=-.9cm,yshift=-2cm]
    \draw[red!65!white] (.5,1.5) to[out=30, in=-30, looseness=4] (.5,.75);

 %   \draw[blue] (.6,1.6) to[out=30, in=-30, looseness=3] (.6,.6)
   % to[out=160, in=160, looseness=4] (.6,.5)
    %to[out=-30, in=30, looseness=2.75] (.6,1.75)
    %to[out=-90-60, in=-90-60, looseness=2.75] (.6,1.6);          
        \end{scope}
%     %bulk
        \begin{scope}[scale=0.75,xshift=-.25cm,yshift=-1.1cm]
     \draw[red!65!white] (3,1.5) ellipse (.75cm and .4cm);
     \draw[blue!65!white] (3,1.5) ellipse (.85cm and .5cm);          
%     %monopoles
     \begin{scope}[xshift=-.5cm,yshift=-0cm]
    \node[cross out,rotate=50,draw=blue!65!white, inner sep=0pt,outer sep=0pt,minimum size=3pt] (a) at (5,1.5) {};
    \node[cross out,rotate=30,draw=blue!65!white, inner sep=0pt,outer sep=0pt,minimum size=3pt] (b) at (5.5,1) {};
    \draw[blue!65!white] (5,1.5) to[out=-30, in=120] (5.5,1);
        \end{scope}
     \end{scope}
\end{scope}
  \end{tikzpicture}
  \caption{Pictorial representation of low energy configurations of
    the $3D$ classical topological paramagnet.  Along red (blue) lines the discrete vorticity of the spins (the gauge flux) is non-zero. In the bulk both of these lines must form closed paths. Energetically they are also encouraged to pair up (middle shape). At a
    boundary (bottom, orange) the flux is zero but
    vorticity lines may end. Since a closed flux loop cannot follow an open vorticity line frustration occurs implying linear confinement of surface vortices. The opposite effect occurs for monopoles of the gauge field
    (crosses) leading again to linear confinement.}
\label{fig:conf}
\end{figure}
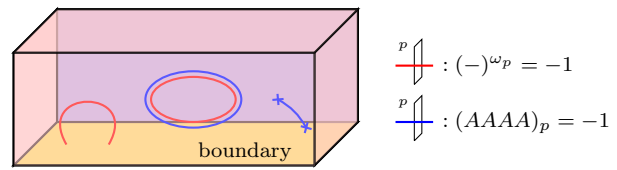

There are two points in phase space where the gauge and spin degrees
of freedom decouple. The trivial case is $J_2 = 0$ which implies free
$\sigma$'s and a standard $\ZZ_2$ gauge theory for the $A$'s. For
$J_1>J_c$, where $J_c=0.762(2)$ is the critical temperature of the
dual Ising model on the cubic lattice, the gauge theory has a perimeter law for Wilson loops and linearly
confines monopoles (open flux lines), but deconfines static charges of the gauge field \cite{Savit}.
%(Recall that the free energy in the presence of two monopoles 
%is given by the partition function for the gauge field where the coupling is reversed along a path 
%connecting the monopoles  ). 
The non-trivial case is
$J_2 > J_c$ and $J_1=0$ and has the same confining bulk physics only
in the composite gauge variable $\tilde{A}$. Notably the transformation
$\tilde{A} = A A_{\sigma}$ can be viewed as acting on the flux degrees
of freedom by multiplying them with vorticity lines. Since vorticity
lines consist of closed polygons, this transformation leaves the
monopole configuration unchanged. Consequently the non-trivial phase
also confines monopoles. See figure \ref{fig:conf} for a representation
of the non-trivial phase.

The above CTP is a robust phase of matter. As in the $2D$ model, the
non-local transformation $\tilde{A}=A A_{\sigma}$ maps local
symmetric and gauge symmetry respecting operators, into local ones,
and leaves the free energy invariant. Respecting these symmetries,
both the monopole confining phases of $\tilde{A}$ and $A$ are well
defined phases \cite{Fradkin1979}. In addition, we found that breaking
the gauge symmetry on an interface or boundary does not destroy the
surface physics (see below) suggesting that gauge symmetry is not crucial here.

\subsection{Surface theory}

To establish the distinction between trivial and non--trivial phases
and to support this nomenclature, we now discuss an interface. For
concreteness we take coordinates $(x,y,z)\in\ZZ^3$ for the vertices of
the lattice and identify the interface as the $x=0$ plane. We also
denote $P_L$ ($P_R$) the plaquettes in the region $x\le 0$ ($x>0$). In
the limit $J_2,J_1\to\infty$, $(AAAA)_{\tilde{p}}=1$ for
$\tilde{p} \in P_R$. By conservation of flux, we find that for all
boundary plaquettes $p \in \partial P$, $(AAAA)_{p}=1$. Consequently
since $J_2$ forces $(-)^{\omega_p}(AAAA)_p=1$, $\omega_p=0$ on the
$2D$ boundary.  The surface partition function in this limit is thus
given by
\begin{align}
  \label{eq:Z_surf}
  Z_{\text{surf},0} &= 
  \sum_{\sigma} \prod_{p\in \partial P}\delta({\omega_p}) =  \sum_{\sigma,\tau} \prod_{p\in \partial P}(\tau)^{\omega_p}\, .
\end{align}

\begin{figure}[h]
  \centering
  \begin{tikzpicture}
  \begin{scope}[decoration={ markings, mark=at position 0.5 with 
      {\arrow[scale=1.5]{>}}}]
    %front
    \draw[postaction={decorate}] (0,0)--(1,0) node[below] {{\tiny $3$}};
    \draw[postaction={decorate}] (1,1)--(1,0) ;
    \draw[postaction={decorate}] (0,1)--(1,1); %node[above] {{\tiny $2$}};
    \draw[postaction={decorate}] (0,1)--(0,0) node[below] {{\tiny $4$}};
    \draw[->] (.5,.75) arc (90:350:.25);
    \node at (-.1,1) {{\tiny $1$}};
    \node at (1.1,1) {{\tiny $2$}};
    %right
    \draw[postaction={decorate}] (1,0)--(1+.5,0.25);
    \draw[postaction={decorate}] (1.5,1+.25)--(1.5,0.25);
    \draw[postaction={decorate}] (1,1)--(1.5,1.25);
    %top
    \draw[postaction={decorate}] (0,1)--(.5,1.25);
    \draw[postaction={decorate}] (.5,1.25)--(1.5,1.25);
	% omega def
	\node at (5,.5) {$\omega_p =
			-\frac{1-\sigma_1\sigma_2}{4}
			-\frac{1-\sigma_2\sigma_3}{4}
			+\frac{1-\sigma_3\sigma_4}{4}
			+\frac{1-\sigma_4\sigma_1}{4}
			$};
  \end{scope}
  \begin{scope}[scale=.7,yshift=-2cm]
	\begin{scope}[xshift=0cm]
		\draw (0,0) rectangle (1,1);
		\node at (.5,-.5) {$0$};
		\node at (0,-.15) {{\tiny $+$}};
		\node at (1,-.15) {{\tiny $+$}};
		\node at (1,1+.15) {{\tiny $+$}};
		\node at (0,1+.15) {{\tiny $+$}};
	\end{scope}
	\begin{scope}[xshift=1.5cm]
		\draw (0,0) rectangle (1,1);
		\node at (.5,-.5) {$0$};
		\node at (0,-.15) {{\tiny $-$}};
		\node at (1,-.15) {{\tiny $+$}};
		\node at (1,1+.15) {{\tiny $-$}};
		\node at (0,1+.15) {{\tiny $+$}};
		\draw[very thick] (0,.5)--(.5,.5)--(.5,1);
		\draw[very thick] (0.5,0)--(.5,.5)--(1,.5);
	\end{scope}
\begin{scope}[xshift=3cm]
		\draw (0,0) rectangle (1,1);
				\node at (.5,-.5) {$0$};
		\node at (0,-.15) {{\tiny $+$}};
		\node at (1,-.15) {{\tiny $+$}};
		\node at (1,1+.15) {{\tiny $+$}};
		\node at (0,1+.15) {{\tiny $-$}};
		\draw[very thick] (0,.5)--(.5,.5)--(.5,1);
	\end{scope}
	\begin{scope}[xshift=4.5cm]
		\draw (0,0) rectangle (1,1);
				\node at (.5,-.5) {$0$};
		\node at (0,-.15) {{\tiny $+$}};
		\node at (1,-.15) {{\tiny $-$}};
		\node at (1,1+.15) {{\tiny $+$}};
		\node at (0,1+.15) {{\tiny $+$}};
		\draw[very thick] (0.5,0)--(.5,.5)--(1,.5);
	\end{scope}
	\begin{scope}[xshift=6cm]
		\draw (0,0) rectangle (1,1);
				\node at (.5,-.5) {$1$};
		\node at (0,-.15) {{\tiny $-$}};
		\node at (1,-.15) {{\tiny $+$}};
		\node at (1,1+.15) {{\tiny $+$}};
		\node at (0,1+.15) {{\tiny $+$}};
		\draw[very thick] (0,.5)--(.5,.5)--(.5,0);
	\end{scope}
	\begin{scope}[xshift=7.5cm]
		\draw (0,0) rectangle (1,1);
				\node at (.5,-.5) {$-1$};
		\node at (0,-.15) {{\tiny $+$}};
		\node at (1,-.15) {{\tiny $+$}};
		\node at (1,1+.15) {{\tiny $-$}};
		\node at (0,1+.15) {{\tiny $+$}};
		\draw[very thick] (0.5,1)--(.5,.5)--(1,.5);
	\end{scope}
	\begin{scope}[xshift=9cm]
		\draw (0,0) rectangle (1,1);
				\node at (.5,-.5) {$0$};
		\node at (0,-.15) {{\tiny $-$}};
		\node at (1,-.15) {{\tiny $+$}};
		\node at (1,1+.15) {{\tiny $+$}};
		\node at (0,1+.15) {{\tiny $-$}};
		\draw[very thick] (0.5,0)--(.5,1);
	\end{scope}
	\begin{scope}[xshift=10.5cm]
		\draw (0,0) rectangle (1,1);
				\node at (.5,-.5) {$0$};
		\node at (0,-.15) {{\tiny $+$}};
		\node at (1,-.15) {{\tiny $+$}};
		\node at (1,1+.15) {{\tiny $-$}};
		\node at (0,1+.15) {{\tiny $-$}};
		\draw[very thick] (0,0.5)--(1,.5);
	\end{scope}
\end{scope}
\end{tikzpicture}
  \caption{(Top) Choice of orientations of links and the formula for
    $\omega_p$ for the front face.  (Bottom) $\sigma$ domain wall
    configurations together with their $\omega_p$ values. Domain wall
    configurations are in bijection with arrow configurations of the
    eight vertex model by associating up/down (right/left) arrows on
    vertical (horizontal) links with presence/absence of thick lines.
  }
  \label{fig:conf_omega0}
\end{figure}
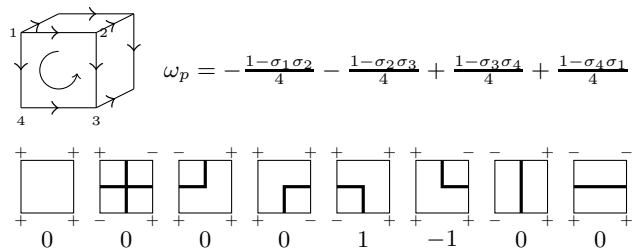

The possible domain wall configurations for $\sigma$'s in $2D$ are
depicted in Fig. \ref{fig:conf_omega0} where a second mapping to arrow
configurations of the eight-vertex model is also discussed. The
constraint $\omega_p=0$ implies a two-in two-out ice rule, supporting
the vorticity interpretation and mapping the surface theory to the critical
six vertex model with an anisotropy parameter
$\Delta = \frac{1}{2}$ \cite{Baxter}.

The latter model model is critical and described by a compact free
boson $\phi$.  This fact can be established with the Coulomb gas method
\cite{Nienhuis1984}, which we now briefly recall.  Denoted by
$S_{\ell}=\pm 1$ the arrow at link $\ell$, note that $S$ is conserved
around a vertex, and one can introduce a height field $h(i)$ on the
same sites where $\sigma$ lives, such that $h$ increases by $\pi$ in
crossing an arrow pointing up from the right.  This discrete height
renormalizes at long distances to a Gaussian free field, a conformal
field theory with central charge $c=1$, and via this mapping one can
compute dimensions of operators.  Noting that
$ \sigma_i\sigma_j = \prod_{\ell\in \Gamma_{ij}} -ie^{i\pi S_\ell /2 }
\propto e^{i h(i)/2 }e^{-i h(j)/2 } $, $\sigma$ is found to have
scaling dimension $3/8$.  Similarly, noting that the two point
function of $\tau$ in eq.~\eqref{eq:Z_surf} corresponds to
inserting two vortices where the height field has discontinuity of
$\pm 4\pi$, $\tau$ has dimension $2/3$.  Identifying $\phi\equiv h/2$,
one has the effective theory
\begin{align}
\mathscr{L}
 &= \frac{g}{4\pi} (\nabla \phi)^2 \, , \quad g=\frac{4}{3}\, .
\end{align}  
The appearance of half integer electric charges follows also naturally
by considering the torus partition function. Indeed on $4L\times 4L'$
lattices, periodic boundary conditions for the $\sigma$'s select only
even frustrations for the height field as it winds around a cycle,
resulting in half integer electric charges and even magnetic charges.
Microscopically, $\sigma$ is a Hermitian linear combination of
$e^{\pm i\phi}$ and $\tau$ of $e^{\pm i\theta}$, $\theta$ being the
dual field.  Therefore, the symmetry is realized as anticipated in the
main text: $\phi \rightarrow \phi + \pi$ and
$\theta \rightarrow \theta + \pi$, as it does in quantum SPTs
\cite{YuanMing2012,Nayak2014}.  We also note that even though the
local weight \eqref{eq:Z_surf} has no such symmetry, the global weight
still has it, due to the global constraint $\prod_p (-1)^{\omega_p}=1$
for a closed manifold.  From this analysis it follows that the lattice
$\ZZ_2\times \ZZ_2$ symmetry is realized in the field theory in an
anomalous chiral way: $\phi \rightarrow \phi + \pi$ and
$\theta \rightarrow \theta + \pi$, where $\theta$ is the dual field.

Let us consider perturbations to this surface model. Adding a
$\sigma\sigma$ term to the boundary action corresponds to the six
vertex model in an external field. Denoted by $H/2$ and $V/2$ the
horizontal and vertical couplings, the theory remains critical within
the region $(e^{2|H|}-1)(e^{2|V|}-1)\le 1$ \cite{Reshetikhin}, the
only effect of $H,V\neq 0$ being renormalizing the stiffness of $\phi$
\cite{Kim}. A ferromagnetic coupling between the $\tau$'s would
generically induce the RG-irrelevant term $\cos(2\theta)$.  
Interestingly, the relevant $\cos(\theta)$ term is forbidden without requiring any fine tuning of the couplings. Formally, it is because of the emergent $\ZZ_2 \times \ZZ_2$ symmetry. Physically, it is because $\pm 2\pi$ vortices are linearly confined by the bulk (see Fig. \ref{fig:conf}). 
Further, a gauge symmetry breaking term ($K\sum_{\ell\in\partial E}A_{\ell}$) can also be studied using duality \cite{Balian} and has no effect on the $\sigma$'s in the limit  $J_2,J_1\to\infty$. 

\subsection{The SPT perspective} 
\label{sec:SPTpersp}

% Lastly, we wish to comment  In Supp.~Mat.~IV. we show that the $3D$ model
% has the same topological response as a $\ZZ_2 \times \ZZ_2$ SPTs
% wherein a fractional symmetry charge to $\pi$ fluxes of a static gauge
% field \cite{Juven2015}.

As discussed in section \ref{sec:AKLT} the $\ZZ_2\times \ZZ_2$
two-dimensional classical topological paramagnet can be related to the
imaginary time partition function of a $1+1D$ quantum SPT phase. In
this section we provide support for the analogous statement in $3D$,
proving that all of the above models are in the same universality
class as the Euclidean time partition function of certain $2+1D$
quantum SPTs.  We will show this by analyzing the responses to gauge
fluxes, or equivalently, the statistical phases obtained by braiding
flux excitations.

As starting point we perform a gauge-to-Ising duality
transformation on the bulk \cite{Balian} trading $A$'s
for spins $\tau$'s on the vertices of the dual lattice, resulting in an equivalent bulk theory with weights:
\begin{align}
  \label{eq:Atau}
  \prod_{p\in P_L}(\tanh J_2)^{\frac{1-\tau_k\tau_l}{2}}
  \prod_{p\in P_R}(\tanh J_1)^{\frac{1-\tau_k\tau_l}{2}}(\tau_k\tau_l)^{\omega_p}\, ,
\end{align}
where $kl$ is the link dual to $p$.  The term
$\prod_{p \in P_R}(\tau_{k}\tau_{l})^{\omega_p}$ is in fact
topological. It is always one in a geometry without interfaces, since
then vorticity lines where $\omega_p=\pm1$ form polygons, and in the
product of $\tau_{k} \tau_{l}$ along each such polygon, each $\tau$
appears an even number of times, and hence the product is always one. Focusing on the analytically tractable case of $J_1=0$ leaves us with the
partition function
\begin{align}
\label{Eq:ModelTau}
Z = \sum_{\tau,\sigma} 
                 \prod_p
                  e^{\tilde{J_2} \tau_{k} \tau_{l}}
  (\tau_{k} \tau_{l})^{\omega_p}\, ,
\end{align} 
where $\tilde{J_2} = \frac{1}{2}\log(\tanh(J_2))$, and here and below
$(kl)$ is the link dual to the plaquette $p$. Since this model now has a $\ZZ_2 \times \ZZ_2$
symmetry, it is natural to seek a quantum counterpart which utilizes
such a symmetry, and these are known as type $ii$ SPT phases \cite{Chen2011,Juven2015,ModularData}.  These SPTs are
characterized by a quantized bulk response to static gauge fluxes.  For a $\ZZ_2 \times \ZZ_2$ symmetry a $\pi$ Ising flux for one symmetry would attract a fractional symmetry charge of the other symmetry. This is the discrete analogue of flux attachment in the integer quantum Hall effect, where a $\pi$ flux would attract half an electron charge \cite{Laughlin1981}. If our model belongs to the same phase as that described by the imaginary
time partition function of one of such $2+1D$ SPTs, it should exhibit
the same flux responses.

We therefore introduce two additional static gauge fields
($B^{\sigma},B^{\tau}$) which are coupled to matter in the usual
manner: we trade each $\tau_{k} \tau_{l}$ with
$\tau_{k} B^{\tau}_{kl} \tau_{l}$ and each $\sigma_{i} \sigma_{j}$
with $\sigma_{i} B^{\sigma}_{i j} \sigma_{j}$. The adjective static
refers to the fact that they are not summed over in the partition
function, which is then:
\begin{align}
  Z(\{B^\tau\},\{B^{\sigma} \})
  &=
    \frac{1}{Z}
  \sum_{\tau,\sigma} 
  \prod_{p}
  e^{\tilde{J_2} \tau_{k}B^\tau_{kl} \tau_{l} }
  (\tau_{k}B_{kl}^\tau \tau_{l})^{\omega_p(B^\sigma)}
  \, ,
\end{align}
where $Z\equiv Z(\{1\},\{1\})$ as above. If we require that both
fluxes are zero everywhere, namely
$\prod_{(ij)\in\partial p} B_{ij}^\sigma=\prod_{(kl)\in\partial p^*} B_{ij}^\tau=1$, 
where $p^*$ is a dual plaquette, we can
rewrite $B_{ij}^\sigma=\tilde{\sigma_i}\tilde{\sigma}_j$, 
 $B_{kl}^\tau=\tilde{\tau_k}\tilde{\tau}_l$, and reabsorb the $B$'s
in the definition of $\sigma,\tau$.  Thus introducing gauge fields
with zero flux is equivalent to set them to $1$. When coupling to
gauge fields, from formula (5) of the main paper the vorticity becomes
\begin{align}
  \label{eq:omegap_B}
  (-)^{\omega_p(B^\sigma)} 
  = 
  \prod_{(ij)\in\partial p}
  \exp\left(i\pi\frac{1-\sigma_iB_{ij}^\sigma \sigma_j}{4} \epsilon^p_{ij}\right)
  \, .
\end{align}
If we now violate the zero flux constraint, then $(-)^{\omega_{p}(B^\sigma)}$
can assume the additional values $\pm i$ on top of $\pm 1$ which it
had before.  A related issue to be discussed is the definition of
plaquette orientations which enter the sign $\epsilon_{ij}^p$.
Changing plaquette orientations corresponds to change the exponent of
\eqref{eq:omegap_B} by an overall sign. For zero $B^{\sigma}$ flux, this
choice is immaterial; however in the case of $\pi$ flux it does matter.  
For definiteness we choose to orient both links and their dual
as the positive direction of the axis of three dimensional space
they are parallel to, and adopt a left-hand rule for defining
clock-wise/anti-close-wise plaquette orientations. 

The topological quantity we wish to calculate concerns the flux
responses in type $ii$ SPT phases with a $\ZZ_2 \times \ZZ_2$ symmetry
and we now recall its definition. Consider then a quantum SPT model
with $\ZZ_2 \times \ZZ_2$ symmetry on a two dimensional lattice, and
denote by $\sigma^{x,z},\tau^{x,z}$ the elementary spin operators, and
by $|\text{gs}\rangle$ its ground state. It can be shown \cite{Juven2015}
that the insertion of a $\pi-$flux associated with one of the
symmetries draws in a fractional symmetry charge associated with the
other symmetry. To probe this we introduce two $B^{\tau}$ $\pi$ fluxes
into the system by creating them and taking them apart at positions
$a,b$. Note that these excitations are string like and a string will
be attached to these two fluxes.  Their worldlines draw a surface
$S_1$ in space time whose interior is swiped by the string. The system
is then let to evolve until it reaches its new ground state, and we
denote the operator that performs this operation by $\pi_{ab}$.
Further, we denote by $S_2$ the set of vertices on a region
surrounding only one of the fluxes and choose this region to be larger
than the correlation length.  

The operator
$\rho_{S_2}=\prod_{i \in S_2} \sigma_{i}^x$ can be interpreted in two ways. First as
creating, evolving and annihilating two $B^\sigma$ $\pi$ fluxes along
the boundary of $S_2$. Second as a measurement of the local Ising charge around just one flux. In a non-trivial type $ii$ SPT with a
$\ZZ_2 \times \ZZ_2$ symmetry, the ratio
$\langle \text{gs} | \pi^{\dagger}_{ab}\rho_{S_2} \pi_{ab} |\text{gs}
\rangle/\langle \text{gs} | \rho_{S_2} \pi^{\dagger}_{ab} \pi_{ab} |\text{gs}
\rangle$ should be equal to $\pm i$ \cite{Juven2015}, the sign depending on which of
the two $B^\tau$ fluxes is encircled by $S_2$.  According to the
previous discussion one can view this as the phase associated with
braiding the two flux excitations (in similar spirit to
Ref. \onlinecite{Levin2012}) or alternatively as a generalization of Laughlin's pumping argument to discrete symmetry
as the $\pi$-flux draws in half an Ising symmetry charge (recall that in this multiplicative notation, an Ising charge is $-1$ and so half a charge is $\pm i$).

\begin{figure}[ht!]
  \begin{tikzpicture}
    \draw[->] (-.5,-.5)--(-.5,1.75) node[left] {$\tau$};
    \begin{scope}
      \node at (1.125,-.5) {(a)};
      \draw[fill=green] (.75,.75) arc (180:0:.75cm and .25 cm);
      \draw[fill=blue,opacity=0.65] (0,0) rectangle (1.5,1.5);
      \draw[fill=green] (.75,.75) arc (180:360:.75cm and .25 cm);
      \node at (.25,1.25) {$S_1$};
      \node at (1.75,.75) {$S_2$};
    \end{scope}
    \begin{scope}[xshift=4cm]
      \node at (1.6,-.5) {(b)};
      \draw[fill=blue,opacity=0.65] (0,0) rectangle (1.5,1.5);
      \node at (.25,1.25) {$S_1$};
      \begin{scope}[xshift=1cm,yshift=0cm]
        \draw[fill=green] (.75,.75) arc (180:0:.75cm and .25 cm);
        \draw[fill=green] (.75,.75) arc (180:360:.75cm and .25 cm);
      \node at (1.5,.75) {$S_2$};
      \end{scope}
    \end{scope}
  \end{tikzpicture}
\caption{Partition function formulation of the generalized Laughlin's
  argument or equivalently the braiding of two $\pi$-fluxes. Across
  the square blue surface $S_1$ the sign of the interaction between two
  $\tau$'s is reversed. Similarly across the oval green surface $S_2$ the
  sign of the interaction between two $\sigma$'s is reversed. The
  ratio between these two partition function equals $\pm i$ for the non-trivial type $ii$ SPT with a $Z_2 \times Z_2$ symmetry.}
\label{Fig:Braiding}
\end{figure}
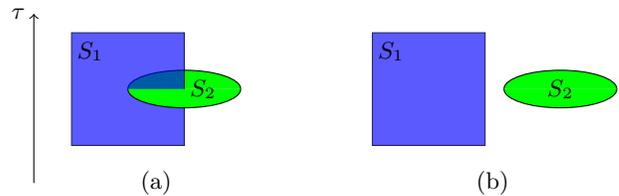

Upon switching to imaginary time, the quantum mechanical overlaps
making up this ratio can be reformulated as partition functions. The
factor
$\langle \text{gs} | \pi^{\dagger}_{ab} \rho_{S_2} \pi_{ab} |\text{gs}
\rangle$ is illustrated in Fig. (\ref{Fig:Braiding}(a)), where across
the $S_1$ surface (blue) the interaction between the $\tau$'s is
reversed and across the $S_2$ surface (green) the interaction between
the $\sigma$'s is reversed. As in the main text, links where the
interaction is reversed are referred to as frustrated. The factor
$\langle \text{gs} | \rho_{S_2} \sigma_{i}^x \pi^{\dagger}_{ab}
\pi_{ab} |\text{gs} \rangle$ illustrated in
Fig. (\ref{Fig:Braiding}(b)) contains the same two elements, however
now these are separated in imaginary time.  More specifically, let us
denote by $G$ and $G^*$ the lattice and its dual, where $\sigma$ and
$\tau$ respectively live. As defined, $S_1$ and $S_2$ will be a
connected region of $G$ and $G^*$ (note the order of $G$ and $G^*$)
across which the $\tau$ and $\sigma$ couplings respectively are
reversed.  By a region here we mean a set of neighbouring plaquettes
and links around them on both the interior and the boundary of the
region. Since it will be clear from the context, we we will write
$(kl) \in S_2$ for links in the region $S_2$.  Further, $\partial S_i$ will
denote the set of links on the boundary of $S_i$. We remark that
frustrated links intersecting $S_1$ ($S_2$) correspond to introducing
a $B^\tau$ ($B^\sigma$) $\pi$ flux on the plaquettes intersecting
$\partial S_1$ $(\partial S_2)$, consistently with the above
discussion.

Before delving into the details of calculating the relevant ratio, let
us give a physical picture supporting why it would come out purely
imaginary. In the presence of non-trivial fluxes, the relation
$\sum_{p\in\text{box}}\omega_p(B^\sigma)=0\mod 2$ does not hold in
general. Instead one has an altered $Z_4$ zero--divergence relation
given by
$2(\sum_{p\in\text{box}}\omega_p(B^\sigma)O^{box}_p)=0\mod 4$, where
$O^{box}_p=1$ ($-1$) if the plaquette's orientation appears as
clock-wise (anti-clock-wise) when viewed from within the box. Using
this new relation one may show that the vorticity line configuration
in the presence of the $B^{\sigma}$ flux loop contains a single
fractional vorticity line encircling $S_2$ as well as other
fluctuating integer vorticity lines. Given the form of the topological
term, the integer vorticity lines cannot contribute imaginary factors
and so we may put them aside for now. Considering the fractional
vorticity line, if it does not cross $S_1$ (case (b)), the term
$\Pi_{(kl) \in \partial S_1} (\tau_k \tau_l)$ is equal to
$1$. Consequently the topological term, which involves a fractional
power of this product, cannot give an imaginary contribution. On the
other hand, if this fractional vorticity line crosses $S_1$ (case (a)), this
product would be $-1$, and the topological term would be purely
imaginary.

We now substantiate the above argument with some simple and exact
computations.  First, notice that there are four cases to consider for
the weight $w(kl)$ per dual link $(kl)$, in case frustrations for both
$\tau$ and $\sigma$ are present:
\begin{align}
  w(kl)
  =
  \begin{cases}
    e^{\tilde{J}_2\tau_k\tau_l}(\tau_k\tau_l)^{\tilde{\omega}_p} & 1): 
    kl \in S_2, \not\cap S_1\\
    e^{-\tilde{J}_2\tau_k\tau_l}(-\tau_k\tau_l)^{\tilde{\omega}_p} & 2): 
    kl \in S_2, \cap S_1\\
    e^{-\tilde{J}_2\tau_k\tau_l}(-\tau_k\tau_l)^{\omega_p} & 3): 
    kl \not\in S_2, \cap S_1\\
    e^{\tilde{J}_2\tau_k\tau_l}(\tau_k\tau_l)^{\omega_p} & 4): 
    kl \not\in S_2, \not\cap S_1
  \end{cases}\, ,
\end{align}
where $\tilde{\omega}_P$ corresponds to $\omega_p(B^\sigma)$ with frustrated
links where $B^\sigma=-1$.
Defined the set of couplings
\begin{align}
  \hat{B}^\tau_\ell &=
  \begin{cases}
    -1 & \ell \cap S_{1}\\
    1 & \ell \not\cap S_{1}
  \end{cases}\, ,\quad
  \hat{B}^\sigma_\ell &=
  \begin{cases}
    -1 & \ell \cap S_{2}\\
    1 & \ell \not\cap S_{2}
  \end{cases}\, ,
\end{align}
the observable of interest is
\begin{align}
  &Z(\{\hat{B}^\tau\},\{\hat{B}^\sigma\}) = \frac{1}{Z}
    \sum \prod_{kl \in S_2, \not\cap S_1} 
    e^{\tilde{J}_2\tau_k\tau_l}(\tau_k\tau_l)^{\tilde{\omega}_p}\\
  &\quad\prod_{kl \in S_2, \cap S_1}
            e^{-\tilde{J}_2\tau_k\tau_l}(-\tau_k\tau_l)^{\tilde{\omega}_p} \\
  &\quad\prod_{kl \not\in S_2, \cap S_1}e^{-\tilde{J}_2\tau_k\tau_l}(-\tau_k\tau_l)^{\omega_p} \\
  & \quad\prod_{kl \not\in S_2, \not\cap S_1}
            e^{\tilde{J}_2\tau_k\tau_l}(\tau_k\tau_l)^{\omega_p} \\
  &=
    \frac{1}{Z}
    \sum \prod_{kl \cap S_1} 
    e^{-\tilde{J}_2\tau_k\tau_l}(-\tau_k\tau_l)^{\omega_p}\\
  &\quad\prod_{kl \not\cap S_1} 
    e^{\tilde{J}_2\tau_k\tau_l}(\tau_k\tau_l)^{\omega_p}
  \prod_{kl \in S_2} 
    (\tau_k\tau_l)^{\tilde{\omega}_p-\omega_p}\\
          &\quad \prod_{kl \in S_2,\cap S_1} 
    (-1)^{\tilde{\omega}_p-\omega_p}\, .
\end{align}
At this point we use the following identity:
\begin{align}
  \label{eq:canceltau}
  \prod_{kl\in S_{2}} (\tau_k\tau_l)^{\tilde{\omega}_p-\omega_p}
  =1 \, .
\end{align}
To prove it, first notice that given the choice of orientation
described in the text above, ${\tilde{\omega}_p-\omega_p}$ gives a
factor $\epsilon_{ij}^p \sigma_i\sigma_j/2$ per frustrated link $ij$.
Then group together all $\tau$'s having a given exponent
$\sigma\sigma'/2$. $\tau$'s appears in pairs for any choice of bond
$\sigma\sigma'$, and cancel either because $\tau^2=1$ or because
$\tau\tau^{-1}=1$.

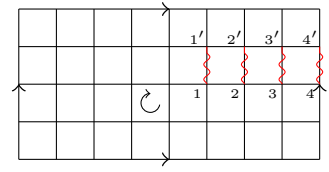
\begin{figure}[ht!]
%\begin{center}
%\includegraphics[scale=0.65]{S1capS2.pdf}
%\includegraphics[width=\columnwidth,clip=true,trim=0 0 0 160]{dis320corr.eps}
%\end{center}
  \begin{tikzpicture}[scale=0.5]
    \begin{scope}[decoration={ markings, mark=at position 0.5 with 
      {\arrow[scale=1.5]{>}}}]
      \draw[postaction={decorate}] (0,0)--(8,0);
      \draw[postaction={decorate}] (8,0)--(8,4);
      \draw[postaction={decorate}] (0,4)--(8,4);
      \draw[postaction={decorate}] (0,0)--(0,4);
    \end{scope}
    \foreach \x in {1,...,7}
    \draw (\x,0)--(\x,4);
    \foreach \y in {1,...,3}  
    \draw (0,\y)--(8,\y);
    \draw[<-] (3.5,1.75) arc (90:350:.25);
    \foreach \x [count=\i] in {5,...,8}
    {
      \draw[red,decorate,
      decoration={snake,amplitude=.4mm,segment length=2mm}]%,post length=1mm}] 
      (\x,2)--(\x,3);
      \node at (\x-.25,2-.25) {{\tiny $\i$}};
      \node at (\x-.25,3+.25) {{\tiny $\i '$}};
    }
  \end{tikzpicture}
\caption{The surface $S_1$. Red bonds are those which intersect $S_2$
  and are frustrated in the $\sigma$ variables.}
\label{fig:S1capS2}
\end{figure}

We now rewrite the partition function in terms of the original $A$
gauge degrees of freedom to take advantage of the change of variables
$A\to \tilde{A}$ as in eq.~\eqref{eq:Atilde}, which decouples gauge and spin
degrees of freedom. Reversing the couplings along
$S_{1}$ for the $\tau$'s corresponds in the $A$ language
 to computing the Wilson loop
along the perimeter of $S_{1}$ (see e.~g.~\cite{Kogut}), so that one
has:
\begin{align}
  &Z(\{\hat{B}^\tau\},\{\hat{B}^\sigma\}) =
  Z^{-1}\sum
  \prod_{p\in S_1}
  (AAAA)_p\\
  &\prod_p e^{J_2 (AAAA)_p (-)^{\omega_p}}
  \prod_{p \in S_1,\cap S_2}(-1)^{\tilde{\omega}_p-\omega_p}\\
  &=
  \left< \prod_{\ell\in\partial S_1}
  \tilde{A}_\ell \right>_{\tilde{A}}
  \left< \prod_{p\in S_1} e^{i\pi\omega_p}
  \prod_{p \in S_1,\cap S_2}e^{i\pi(\tilde{\omega}_p-\omega_p)}\right>_{\sigma}\, 
\end{align}
where the average $\langle ... \rangle_{\tilde{A}}$ is taken with the partition function of $\tilde{A}$'s alone, and the average $\langle ... \rangle_{\sigma}$ is taken with the trivial partition function for the $\sigma$'s that gives a weight of $1$ to each $\sigma$ configuration. 
The last term in the $\sigma$ expectation values involves the links illustrated in figure 
\ref{fig:S1capS2}.

Due to cancellations on the internal edges, now we have the following
identities -- recall also the discussion around \eqref{eq:canceltau}, and 
use the notation
of sites along the frustrations as in fig.~\ref{fig:S1capS2}:
\begin{align}
  &\prod_{p\in S_1} e^{i\pi\omega_p}
  =
  \prod_{(ij)\in \partial S_1} i^{\epsilon(ij)^p \frac{1-\sigma_i\sigma_j}{2}}\, ,\\
  &\prod_{p \in S_1,\cap S_2}e^{i\pi(\tilde{\omega}_p-\omega_p)}
  = 1 \text{ if (b) : } S_1 \cap S_2=\emptyset\\
  &\prod_{p \in S_1,\cap S_2}e^{i\pi(\tilde{\omega}_p-\omega_p)}
  =\\
  &%\!\!\!\!\!\!\!\!\!\!\!\!\!\!\!\!\!\!\!\!\!\!\!\!\!\!\!
    e^{i\frac{\pi}{2}
  (-\sigma_1\sigma_1'+\sigma_1\sigma_1'-\sigma_2\sigma_2'+\sigma_3\sigma_3'
  -\sigma_3\sigma_3'+\sigma_4\sigma_4')}\\
  &=e^{i\frac{\pi}{2}\sigma_4\sigma_4'} 
    \text{ if (a) : } S_1 \cap S_2\not=\emptyset
  \, .
\end{align}
Therefore, in both (a),(b) cases the $\sigma$ expectation value
reduces to a one dimensional classical spin chain along
$\partial S_{1}$ which can be easily solved via transfer matrix.  The
presence of frustration in case (a) corresponds to introducing a twist
by the matrix $ e^{i\frac{\pi}{2}\sigma\sigma'}$.  Under the
assumption of a rectangular perimeter $\partial S_1$ of length $2N$,
with the branching structure as in fig.~\ref{fig:S1capS2}, the
$\sigma$ expectation value in the (a) case is (setting
$\sigma_{2N+1}\equiv \sigma_{1}$):
\begin{align}
  &\left< \prod_{p\in S_1} e^{i\pi\omega_p}
  \prod_{p \in S_1,\cap S_2}e^{i\pi(\tilde{\omega}_p-\omega_p)}\right>_{\sigma}=\\
  &= 2^{-|\partial S_1|}
  \Tr \left[
  \begin{pmatrix}
    i&-i\\
    -i&i
  \end{pmatrix}
  \begin{pmatrix}
    1&i\\
    i&1
  \end{pmatrix}^N
  \begin{pmatrix}
    1&-i\\
    -i&1
  \end{pmatrix}^N \right]\\
  &=
  i 2^{1-N}\, ,
\end{align}
Let us remark that the problem has a chirality given by the branching
structure. If $S_2$ crossed $S_1$ on the left boundary instead of on the
right, the twist matrix would have been $e^{-i\frac{\pi}{2}\sigma\sigma'}$, and
it would have produced an extra minus sign. 
If the flux arrangement is as in Fig. \ref{Fig:Braiding} (b), the only difference in the result is
the absence of the twist matrix appearing first in the above trace. The sole net effect of this is to remove the $i$ factor and therefore the desired ratio is
\begin{align}
  Z^{\text{(a)}}/Z^{\text{(b)}} = \pm i\, ,
\end{align}
depending if $S_2$ crosses $S_1$ on its right ($+$) or left ($-$).
We have thus shown that our model has the same response to $\pi$
fluxes as the related quantum SPT phase.

\subsection{Generalizations}

As done in section \ref{sec:gen_2d} for the $2D$ case, we now sketch
generalisations of the $3$D model beyond the case of a $\ZZ_2$ symmetry.

\subsubsection{Discrete vorticity and cellular cohomology}
\label{sec:cohom}

We first address the mathematical description of the discrete
vorticity in terms of cellular cohomology
which allow for its generalization.  We will then outline a
classification of CTPs within this framework and analyze some specific
models.

Simplicial and cellular cohomology are toolboxes used lattice gauge
theories (See e.g.~ \cite{Cohomology1991}.).  The first requires us to
work strictly with simplexes while the second permits more general
types of cells, in particular the cubic lattice. Let us quickly
describe the necessary mathematical details. A reader interested only
the generalized definition of the discrete vorticity for $G=Z_N$ may
skip directly to Eq. \ref{Eq:DiscreteVorticityZN}.

We denote the sets of sites, edges, plaquettes and boxes of the cubic
lattice by $V,E,P,B$ respectively, and call their elements
alternatively $0$-,$1$-,$2$- and $3$-cells. In the obvious manner each
of these sets describes the boundary of the latter one. The relations
between cells and their boundaries can be captured in several ways:
One is using incidence numbers, where $[a:b]$, with $a$ a $d$-cell and
$b$ a $d+1$ cell.  These take three possible integer values, $-1,0,1$,
which satisfy sum rules, such as
$\sum_{e\in E} [v:e][e:p] = 0, \sum_{p\in P}[e:p][p:b]=0$.
Alternatively, one can simply orient the edges and plaquettes and then
$[v:e]$ will be $0,1$ or $-1$ is $v$ is not a boundary of $e$, $v$
is at the end of $e$ or $v$ is at the beginning of $e$. Similarly
$[e:p]$ is $0$ if $e$ is not an edge of $p$, $1$ if $e$ is aligned
along the orientation of $p$ or $-1$ if it is opposite. One can easily
verify that these definitions satisfy the sum rules.

Below we use $i,j,k,..$ for vertex indices, $\epsilon_{ij}= 1$ ($-1$) if the edge $ij$ is oriented from $i$ to $j$ ($j$ to $i$) and $\epsilon^p_{ij}= 1$ ($-1$) if the edge $ij$ is oriented along the orientation of the plaquette (against it). 

% To facilitate comparison with mathematical literature on cellular cohomology, it is worth mentioning a third approach for defining incident numbers based on CW-complexes. Here the relation between an n-cells ($e$) and its boundary is captured by a map ($\phi_e$) from the n-disk ($D_n$) and the $n-1$-sphere ($S_{n-1}$) onto the inner part and the boundary of $e$ respectively. On the boundary this map is surjective while inside the cell it is an isomorphism. In this formalism, and focusing on 2-cells for simplicity $[e_E:e_p]$ is defined as followed: First one identifies all the vertices as a single base point ($E/V$) . This causes each edge to appear as a circle and $E/V$ becomes a union of circles all touching at this base point. There are now two maps from $S_1$ to $E/V$ to consider. The first is an isomorphism coming from $\phi_{e_E} : D_1 \rightarrow e_E$, along with the identification of vertices. The second subjective mapping comes from  $\phi_{e_p}$, when restricted to its $S_1$ boundary. Together they define a mapping a surjective mapping from $S_1$ to a union of oriented circles. The degree of this map with respect to each of the circles in the union (i.e. the edges) is the incident number. 

To define a cellular cohomology structure (or physically a gauge theory coupled to matter)
the following steps are needed: First we pick an abelian group (the
gauge group) $G$ and call an assignment $g : V \to G$ a
$0$-cochain (matter field), $A : E \to G$ a $1$-cochain (gauge field),
and $F : E \to G$ a $2$-cochain (curvature/flux field). We denote the set
of $d$-cochains by $C^d$. The coboundary operator $\delta$ (see Ref.~\onlinecite{Cohomology1991}) maps $C^d$
to $C^{d+1}$, and is nilpotent, $\delta^2=0$. In particular,
$(\delta g)_{ ij \in E} = g_i g^{-1}_j$, where the order of $ij$
is chosen according to the orientation of the edge, is the trivial
1-cocycle. (If $G$ is a generic abelian group we will use the notation
$(\delta g)_{ij} = g_i - g_j$, and if $G=\mathbb{Z}_2$,
$g_i = (1-\sigma_i)/2$, where $\sigma_i=\pm 1$ is the variable used in
the main text.).  In general, given $\alpha\in C^d$,
$\beta=\delta\alpha$ is a trivial $d+1$-cochain, and if $\beta=0$,
then $\alpha$ is called a $d$-cocycle.  Next, one can define an
equivalence relation where two $d$-cocycles are equivalent if their
differ by a trivial $d$-cochain: $\alpha_1-\alpha_2=\delta\gamma$,
with $\gamma\in C^{d-1}$.  The equivalence classes of $d$-cocycles
then obey a group structure known as the $d$ cohomology
group $H^d(G)$.

We consider now an exact sequence of abelian groups of the type
\begin{align}
  \label{eq:exact_seq}
  0 \rightarrow G \overset{f}{\rightarrow} \tilde{G} \overset{h}{\rightarrow} G \rightarrow 0\, ,
\end{align}
and construct the map $B=f^{-1} \delta h^{-1}$, which is applied to
a trivial $1$-cocycle $\delta g$ to produce a $2$-cocycle.  The map $B$
is called a Bockstein homomorphism
\cite{Hatcher2002,Kapustin2014} and is well-defined
given $h^{-1},f^{-1}$. Further, it maps $d$-cocycles to
$d+1$-cocycles and introduces a homomorphism between $H^d(G)$ and
$H^{d+1}(G)$. In physical terms, it maps a matter configuration to gauge flux configurations with no monopoles.

In general, there is a variety of exact sequences one can consider and
hence a variety of Bockstein homomorphisms. These can be classified by
classifying the exact sequences upon which their are based. Short
exact sequences of the form \eqref{eq:exact_seq} involving abelian
groups are equivalent to central extension of $G$ by $G$
(s.t.~$G=\tilde{G}/G$). The trivial extension is defined by
$\tilde{G}=G\times G,f(a) = (a, 0)$ and $h((a, b)) = b$. Non-trivial
extensions are classified by the second group cohomology 
$H^2(G,G)$. For $G = Z_N$ with $N$ prime, one finds that
$H^2(Z_N,Z_N) = Z_N$ and so $N$ distinct choices of discrete vorticity
exist.

If we specify to $G=\mathbb{Z}_2, \tilde{G}=\mathbb{Z}_4$, and
$f(a)=2 a, h(a)=a\mod 2$, the Bockstein homomorphism $B$ produces
precisely $\omega_p \mod 2$ and the 2-cocycle condition implies zero
divergence. Moreover, since $B$ is a homomorphism and
$\delta g$ is a trivial $1$-cocycle, the 2-cocycle must be trivial as
well and hence there exists a $1$-cochain (a gauge field, $A$) such
that $\delta A = \omega_p$.

We can now use $B$ to define discrete vorticities for other abelian
groups. Consider for instance the case $G=\mathbb{Z}_N$, $N$ prime,
$\tilde{G}=Z_{N^2}$, and:
\begin{align}
  f(a) &= N a \, ,&  h_\ell(a) &= \ell a \mod N\, , &
  \ell &= 0,1,\dots,N-1\, .
\end{align}
Each choice of $\ell$ realizes one of the $N$ nonequivalent central
extensions of $\ZZ_N$ by $\ZZ_N$, and leads to a different Bockstein
homomorphism with $\ell=0$ being the trivial case.
Setting $B_\ell = f^{-1} \delta h^{-1}_\ell$ yields a discrete vorticity
generalising eq.~\eqref{eq:omegap}:
\begin{align}
\label{Eq:DiscreteVorticityZN}
  \omega_p^{(\ell)} = 
  \frac{1}{N}
  \sum_{(ij)\in \partial p}\epsilon_{ij}^p\ell \left( g_i-g_j\right)\mod N^2
  \, .
\end{align}
where $i$ and $j$ in the above are chosen such that $i$ ($j$) is at
the start (end) of the edge $(\vec{ij})$ and $\epsilon_{ij}^p= 1$ ($-1$)
if the edge is oriented with (against) the plaquette $p$.
(Equivalently $\epsilon^p_{ij}$ is the incidence number
$[(\vec{ij}):p]$ in the notation of Ref. ~\onlinecite{Cohomology1991}.)
Explicitly, referring to figure \ref{fig:conf_omega0}, it reads:
\begin{equation}
  \begin{split}
\omega_p^{(\ell)} =
&\frac{1}{N}
\Big(
\ell
\big(
    -(g_1-g_2)
    -(g_2-g_3)\\
    &+(g_4-g_3)
    +(g_1-g_4)
  \big)
  \mod N^2\Big)
  \, .
  \end{split}
\end{equation}
The non-triviality of this expression is due to the fact that
the terms $(g_i-g_j)$ are understood in $\ZZ_N$.

Lastly we comment on the connection between the above cellular-cohomology approach and the
group-cohomology approach to SPTs \cite{Chen2011}. Quantum SPTs at $d+1$ spatial
dimensions with a symmetry $Q$ are classified by the group-cohomology group $H^{d+1}(Q,U(1))$. In our classical
context $d+1$ is actually the overall dimension, and so one may expect
that our phase is contained in $H^3(Q,U(1))$. If our matter fields
posses a $Z_N$ symmetry and the gauge symmetry is $Z_N$, the relevant
symmetry group in our context is $Q=Z_N \times Z_N$. (This is shown
explicitly in the next section for $N=2$.)  Considering 
$Q=Z_N \times Z_N$, the Kunneth formula \cite{chen2014symmetry} tells
us that $H^3(Z_N \times Z_N,U(1))=\ZZ_N^3$ contains
$H^2(Z_N,H^1(Z_N,U(1))) = H^2(Z_N,Z_N)$ which is also the quantity
which classifies central extensions, as discussed above. It would be
interesting to find the exact correspondence between
$H^3(G\times G',U(1))$ and possible CTPs. In particular find out
whether every element in $H^3(G \times G',U(1))$ corresponds to a
classical (or local sign free) partition function.

\subsubsection{Discrete vorticity models of 3D CTPs with $G=G'=\ZZ_N$}
Using the above definition of a discrete vorticity for $G=Z_N$ one can
readily define more general models of 3D CTPs. To this end we consider
a cubic lattice with vertices indexed by $i$, oriented edges pointing
from $i$ to $j$ by $(ij)$ and oriented plaquettes indexed by
$p$. The model has $\sigma_i \in \ZZ_N$ degrees of freedom on vertices
and $A_{ij} \in \ZZ_N$ degrees of freedom on edges of the
lattice. As in the two-dimensional case $\ZZ_N$ degrees of freedom
take values in the roots of unity ($e^{2\pi i \alpha /N}$). (However we
still represent $\omega^{(\ell)}_p$ as a number between $0,\dots,N-1$). In
this notation the generalized model is given by
\begin{align}
  \label{eq:E_3D_ZN}
  -\beta\mathscr{H}
  &= \sum_p  J_{\ell'} e^{\frac{2\pi i \omega^{(\ell')}_p}{N}} (AAAA)_p  + c.c. \\ \nonumber 
  &+ \sum_p J_{\ell} e^{\frac{2\pi i \omega^{(\ell)}_p}{N}} (AAAA)_p  + c.c. \, ,
\end{align}
with $\omega^{(\ell)}_p$ being the discrete vorticity from Eq. (\ref{Eq:DiscreteVorticityZN}), which depends on $g_i$ defined by $\sigma_i = e^{2\pi i g_i /N}$ and $(AAAA)_p \in \ZZ_N$ is the product of $A^{\epsilon^p_{ij}}_{ij}$'s along the plaquette $p$. 

First let us analyze the case when only $J_{\ell}$ is non-zero. The previous discussion on $\omega^{(\ell)}_p$ shows that for every $\sigma$ configuration there is a $A_{\sigma}$ configuration such that $(A_{\sigma}A_{\sigma}A_{\sigma}A_{\sigma})_p = \omega^{(\ell)}_p$. Thus going to the composite gauge variable $\tilde{A} = A A_{\sigma}$ one obtains $-\beta\mathscr{H} = J_{\ell} (\tilde{A} \tilde{A} \tilde{A} \tilde{A})_p$--- a pure $\ZZ_N$ lattice gauge theory. 

Performing a generalized Kramers--Wannier duality  \cite{Balian} a $\ZZ_N$ lattice gauge theory becomes a $3D$ clock model with rotor variables taking values in $\ZZ_N$. For prime $N$, so that $\ZZ_N$ doesn't have any subgroups, the model will exhibit two distinct thermodynamic phases: A disordered phase where the rotors are disordered and an ordered phase of the rotors separated by a second order phase transition at $J_c$. In gauge theory terms,  these correspond respectively to a phase with short flux loops ($J_{\ell}>J_{c}$) and one with large flux loops ($J_{\ell}<J_{c}$). Following the exact same reasoning as done for the $Z_2$ case, we find that the former phase confines defects of the constraint and since $\sigma$ can fluctuate freely, it clearly doesn't break any symmetry. Consequently it is an admissible phase in our classification.

We argue that the phase obtained for $J_{\ell}>J_{c}$ is a
classical topological phase of type $\ell$. As discussed previously,
it is a phase since local symmetry and gauge respecting perturbation
in the $\sigma,A$ degrees of freedom map to local symmetry and gauge
respecting perturbation in the $\sigma,\tilde{A}$ notation and
vice-versa. Knowing that the latter is a well defined thermodynamic
phase then implies that the former one is well defined as well. To see
why different $\ell$ correspond to distinct phases let us consider an
interface between a phase with large
$J_{\ell} \rightarrow \infty,J_{\ell'}=0$ on the left and $J_{\ell'} \rightarrow \infty, J_{\ell} = 0$
on the right.  At the interface,
$\omega^{(\ell)}_p=\omega^{(\ell')}_p$.  Now, since
$\omega^{(\ell)}_p=\ell \omega_p^{(1)} \mod N$ and $N$ is prime,
consistency implies either $\ell=\ell'$ or $\omega_p^{(1)}=0$.
Supposing $\ell\neq \ell'$, this shows that just as in the $Z_2$ case,
the boundary is described by a $2D$ statistical mechanical model where
a zero vorticity constraint is imposed on every square.  Taking
$J_{\ell} > J_c$ but finite on the left and $J_{\ell'} > J_c$ on the right, will result in a
physically similar scenario where flux lines crossing the interface
are confined to neutral pairs by the bulks.  We will argue momentarily
that the model with zero vorticity is gapless.  This, together with
the relations to the group cohomology classification of the previous
section, strongly suggests that different $\ell$ correspond to
different phases. One way of proving this would be to generalize the
arguments of section \ref{sec:SPTpersp} to $\ZZ_N$, and is left for
future work.

Let us analyze the resulting theory on the two dimensional interface.
We first count the number of zero vorticity constraints at a
plaquette.  We change variables from site to links variables
$s_{ij}=g_i-g_j$, where as before $\sigma_i=e^{2\pi i g_i/N}$.  The four link variables
around a plaquette can assume only $N^3$ since a global shift of $g_i$
leaves the link variables unchanged.  (In the following we will ignore
the multiplicative factor $N$ in the weight produced by this change of
variables.)  For the purpose of counting the zero vorticity
configurations, we can ignore this constraint and consider the link
variables independent since the missing $N$ configurations have
non-zero vorticity.  We are thus left with a vertex model, where each
link has $N$ states and zero vorticity becomes an interaction at
vertices of dual lattice.  Further, the zero vorticity constraint is
the same for any $\ell$ in \eqref{Eq:DiscreteVorticityZN} and
w.r.t.~the labelings of vertices and orientations as in figure
\ref{fig:conf_omega0}, it reads:
\begin{align}
  -s_{12} - s_{23} + s_{43} + s_{14} = 0\, .
\end{align}
If the $N$ states are labeled $-S,\dots,S$, with $S=\frac{N-1}{2}$,
this coincides with U$(1)$ invariant configurations of spin-$S$ vertex
models, and the resulting number of non-zero configurations is
\begin{align}
  \frac{N}{3} (2 N^2 + 1) =
  6, 19, 44, 85, \dots
\end{align}
Apart from the already discussed $N=2$ case, other values of $N$ may
not correspond to integrable weights for the vertex model, as we will
discuss now for the case $N=3$, where the number of vertices is $19$.
In such case, there are two classes of integrable $19$ vertex models,
both of which can be related to a loop model, see
e.g.~\cite{YUNG1995}. In particular, our model gives uniform
weight one to each vertex and cannot be related to a loop model, at
least not in the standard fashion where states of labels $\pm 1$ are
associated to oriented strands of loops and states of labels $0$ to
vacancies. Nonetheless, this model belongs to a class of models
studied numerically in relation with Berezinskii-Kosterlitz-Thouless transition in
\cite{Honda1997}, suggesting that the model is critical and with $c=1$.

\section{Conclusion}
In this work we have introduced a topological classification scheme of classical statistical mechanical systems. This involved defining the objects of the classification (admissible phases), the equivalence relations between them (continuous deformation without phase transitions) and lastly showing that the classification is not trivial by giving concrete examples of admissible phases which are inequivalent. We have found $N$ distinct models for CTPs in $2D$ and $3D$ for systems with a $\ZZ_N$ symmetry and defects carrying a $\ZZ_N$ charge. An important question concerning the ability to identify the topological index or equivalence class given the bulk behavior of a particular model is left for future work. 

The CTPs introduced in this work, together with the ones discussed in
\onlinecite{Geraedts2013,You2016}, describe, to the best of our
knowledge, novel types of topological classical phases of matter. The models given here are, arguably, the simplest and most minimal ones having just a spin degree of freedom per site and per link. Another salient feature is that they can be simulated using classical
Monte-Carlo. They may thus serve as a test-bed for studying
various open questions concerning both classical topological
paramagnets and their quantum counterparts \cite{Chen2011}. These
concern the nature of phase transition between trivial and non-trivial
phases \cite{You2016}, the effect of disorder on the surfaces and on
phase transitions, and the precise implications of the bulk-boundary
correspondence \cite{Scaffidi2016}.

It would be highly desirable to find possible experimental
realizations of such CTPs. In the field of
quantum bosonic SPTs \cite{Chen2011}, experimental realizations are so
far limited to $1+1D$ \cite{Buyers1986}.
Being free from the stringent requirement of quantum coherence, and
based on simple microscopic ingredients, the classical counterparts
introduced here may prove easier to realize. Indeed similar classical
systems, such as artificial spin-ice systems, have been successfully
realized \cite{Wang2006,Cumings2008,Roderich2013} using ferromagnetic wires as well as tiling molecules \cite{Blunt}. 
The $2D$ model we discussed could potentially be realized from the same
microscopic ingredients. 

Finally, it would be interesting to further explore the classification
question we propose in this work. For instance by considering other types of symmetries and constraints. 
Certainly there should be some relation with the group
cohomology classification of bosonic SPTs with a trivial bulk
\cite{Chen2011} however it may not be one to one. Indeed some SPTs may
suffer from sign problems in Monte-Carlo while others do
not. Conversely, it may be that enforcing hard constraints or gauge symmetries allows for
new types of quantum phases. Indeed hard constraints in classical
systems may result in a genus dependent ergodicity breaking
\cite{Moessner2001,PhysRevB.93.205112} whereas genus dependent ground state degeneracy is
not part of the cohomology classification of
Ref. (\onlinecite{Chen2011}).

We are grateful to P.~Fendley, T.~Scaffidi and S.H.~Simon for
stimulating discussions. Z.R. was supported by the
European Union's Horizon 2020 research and innovation programme under
the Marie Sklodowska-Curie grant agreement No. 657111.  R.B. was
supported by the EPSRC Grants EP/I031014/1 and EP/N01930X/1.  Both
authors contributed equally to this work.

\bibliography{CTP}

%merlin.mbs apsrev4-1.bst 2010-07-25 4.21a (PWD, AO, DPC) hacked
%Control: key (0)
%Control: author (8) initials jnrlst
%Control: editor formatted (1) identically to author
%Control: production of article title (-1) disabled
%Control: page (0) single
%Control: year (1) truncated
%Control: production of eprint (0) enabled
\begin{thebibliography}{55}%
\makeatletter
\providecommand \@ifxundefined [1]{%
 \@ifx{#1\undefined}
}%
\providecommand \@ifnum [1]{%
 \ifnum #1\expandafter \@firstoftwo
 \else \expandafter \@secondoftwo
 \fi
}%
\providecommand \@ifx [1]{%
 \ifx #1\expandafter \@firstoftwo
 \else \expandafter \@secondoftwo
 \fi
}%
\providecommand \natexlab [1]{#1}%
\providecommand \enquote  [1]{``#1''}%
\providecommand \bibnamefont  [1]{#1}%
\providecommand \bibfnamefont [1]{#1}%
\providecommand \citenamefont [1]{#1}%
\providecommand \href@noop [0]{\@secondoftwo}%
\providecommand \href [0]{\begingroup \@sanitize@url \@href}%
\providecommand \@href[1]{\@@startlink{#1}\@@href}%
\providecommand \@@href[1]{\endgroup#1\@@endlink}%
\providecommand \@sanitize@url [0]{\catcode `\\12\catcode `\$12\catcode
  `\&12\catcode `\#12\catcode `\^12\catcode `\_12\catcode `\%12\relax}%
\providecommand \@@startlink[1]{}%
\providecommand \@@endlink[0]{}%
\providecommand \url  [0]{\begingroup\@sanitize@url \@url }%
\providecommand \@url [1]{\endgroup\@href {#1}{\urlprefix }}%
\providecommand \urlprefix  [0]{URL }%
\providecommand \Eprint [0]{\href }%
\providecommand \doibase [0]{http://dx.doi.org/}%
\providecommand \selectlanguage [0]{\@gobble}%
\providecommand \bibinfo  [0]{\@secondoftwo}%
\providecommand \bibfield  [0]{\@secondoftwo}%
\providecommand \translation [1]{[#1]}%
\providecommand \BibitemOpen [0]{}%
\providecommand \bibitemStop [0]{}%
\providecommand \bibitemNoStop [0]{.\EOS\space}%
\providecommand \EOS [0]{\spacefactor3000\relax}%
\providecommand \BibitemShut  [1]{\csname bibitem#1\endcsname}%
\let\auto@bib@innerbib\@empty
%</preamble>
\bibitem [{\citenamefont {Hasan}\ and\ \citenamefont {Kane}(2010)}]{Hasan2010}%
  \BibitemOpen
  \bibfield  {author} {\bibinfo {author} {\bibfnamefont {M.~Z.}\ \bibnamefont
  {Hasan}}\ and\ \bibinfo {author} {\bibfnamefont {C.~L.}\ \bibnamefont
  {Kane}},\ }\href@noop {} {\bibfield  {journal} {\bibinfo  {journal} {Reviews
  of Modern Physics}\ }\textbf {\bibinfo {volume} {82}},\ \bibinfo {pages}
  {3045} (\bibinfo {year} {2010})}\BibitemShut {NoStop}%
\bibitem [{\citenamefont {Ilan}\ \emph {et~al.}(2015)\citenamefont {Ilan},
  \citenamefont {de~Juan},\ and\ \citenamefont {Moore}}]{Ilan2014}%
  \BibitemOpen
  \bibfield  {author} {\bibinfo {author} {\bibfnamefont {R.}~\bibnamefont
  {Ilan}}, \bibinfo {author} {\bibfnamefont {F.}~\bibnamefont {de~Juan}}, \
  and\ \bibinfo {author} {\bibfnamefont {J.~E.}\ \bibnamefont {Moore}},\ }\href
  {\doibase 10.1103/PhysRevLett.115.096802} {\bibfield  {journal} {\bibinfo
  {journal} {Phys. Rev. Lett.}\ }\textbf {\bibinfo {volume} {115}},\ \bibinfo
  {pages} {096802} (\bibinfo {year} {2015})}\BibitemShut {NoStop}%
\bibitem [{\citenamefont {Wu}\ \emph {et~al.}(2011)\citenamefont {Wu},
  \citenamefont {Peeters},\ and\ \citenamefont {Chang}}]{Wu2011}%
  \BibitemOpen
  \bibfield  {author} {\bibinfo {author} {\bibfnamefont {Z.}~\bibnamefont
  {Wu}}, \bibinfo {author} {\bibfnamefont {F.~M.}\ \bibnamefont {Peeters}}, \
  and\ \bibinfo {author} {\bibfnamefont {K.}~\bibnamefont {Chang}},\ }\href
  {\doibase http://dx.doi.org/10.1063/1.3581887} {\bibfield  {journal}
  {\bibinfo  {journal} {Applied Physics Letters}\ }\textbf {\bibinfo {volume}
  {98}},\ \bibinfo {eid} {162101} (\bibinfo {year} {2011})}\BibitemShut
  {NoStop}%
\bibitem [{\citenamefont {Ojeda-Aristizabal}\ \emph {et~al.}(2012)\citenamefont
  {Ojeda-Aristizabal}, \citenamefont {Fuhrer}, \citenamefont {Butch},
  \citenamefont {Paglione},\ and\ \citenamefont {Appelbaum}}]{Ojeda2012}%
  \BibitemOpen
  \bibfield  {author} {\bibinfo {author} {\bibfnamefont {C.}~\bibnamefont
  {Ojeda-Aristizabal}}, \bibinfo {author} {\bibfnamefont {M.~S.}\ \bibnamefont
  {Fuhrer}}, \bibinfo {author} {\bibfnamefont {N.~P.}\ \bibnamefont {Butch}},
  \bibinfo {author} {\bibfnamefont {J.}~\bibnamefont {Paglione}}, \ and\
  \bibinfo {author} {\bibfnamefont {I.}~\bibnamefont {Appelbaum}},\ }\href
  {\doibase http://dx.doi.org/10.1063/1.4733388} {\bibfield  {journal}
  {\bibinfo  {journal} {Applied Physics Letters}\ }\textbf {\bibinfo {volume}
  {101}},\ \bibinfo {eid} {023102} (\bibinfo {year} {2012})}\BibitemShut
  {NoStop}%
\bibitem [{\citenamefont {Fu}\ and\ \citenamefont {Kane}(2008)}]{FuKane2008}%
  \BibitemOpen
  \bibfield  {author} {\bibinfo {author} {\bibfnamefont {L.}~\bibnamefont
  {Fu}}\ and\ \bibinfo {author} {\bibfnamefont {C.~L.}\ \bibnamefont {Kane}},\
  }\href {\doibase 10.1103/PhysRevLett.100.096407} {\bibfield  {journal}
  {\bibinfo  {journal} {Phys. Rev. Lett.}\ }\textbf {\bibinfo {volume} {100}},\
  \bibinfo {pages} {096407} (\bibinfo {year} {2008})}\BibitemShut {NoStop}%
\bibitem [{\citenamefont {{Ryu}}\ \emph {et~al.}(2010)\citenamefont {{Ryu}},
  \citenamefont {{Schnyder}}, \citenamefont {{Furusaki}},\ and\ \citenamefont
  {{Ludwig}}}]{TenFold2010}%
  \BibitemOpen
  \bibfield  {author} {\bibinfo {author} {\bibfnamefont {S.}~\bibnamefont
  {{Ryu}}}, \bibinfo {author} {\bibfnamefont {A.~P.}\ \bibnamefont
  {{Schnyder}}}, \bibinfo {author} {\bibfnamefont {A.}~\bibnamefont
  {{Furusaki}}}, \ and\ \bibinfo {author} {\bibfnamefont {A.~W.~W.}\
  \bibnamefont {{Ludwig}}},\ }\href {\doibase 10.1088/1367-2630/12/6/065010}
  {\bibfield  {journal} {\bibinfo  {journal} {New Journal of Physics}\ }\textbf
  {\bibinfo {volume} {12}},\ \bibinfo {eid} {065010} (\bibinfo {year}
  {2010})},\ \Eprint {http://arxiv.org/abs/0912.2157} {arXiv:0912.2157
  [cond-mat.mes-hall]} \BibitemShut {NoStop}%
\bibitem [{\citenamefont {Qi}(2011)}]{Qi2011}%
  \BibitemOpen
  \bibfield  {author} {\bibinfo {author} {\bibfnamefont {X.-L.}\ \bibnamefont
  {Qi}},\ }\href {\doibase 10.1103/PhysRevLett.107.126803} {\bibfield
  {journal} {\bibinfo  {journal} {Phys. Rev. Lett.}\ }\textbf {\bibinfo
  {volume} {107}},\ \bibinfo {pages} {126803} (\bibinfo {year}
  {2011})}\BibitemShut {NoStop}%
\bibitem [{\citenamefont {Chen}\ \emph {et~al.}(2013)\citenamefont {Chen},
  \citenamefont {Gu}, \citenamefont {Liu},\ and\ \citenamefont
  {Wen}}]{Chen2011}%
  \BibitemOpen
  \bibfield  {author} {\bibinfo {author} {\bibfnamefont {X.}~\bibnamefont
  {Chen}}, \bibinfo {author} {\bibfnamefont {Z.-C.}\ \bibnamefont {Gu}},
  \bibinfo {author} {\bibfnamefont {Z.-X.}\ \bibnamefont {Liu}}, \ and\
  \bibinfo {author} {\bibfnamefont {X.-G.}\ \bibnamefont {Wen}},\ }\href
  {\doibase 10.1103/PhysRevB.87.155114} {\bibfield  {journal} {\bibinfo
  {journal} {Phys. Rev. B}\ }\textbf {\bibinfo {volume} {87}},\ \bibinfo
  {pages} {155114} (\bibinfo {year} {2013})}\BibitemShut {NoStop}%
\bibitem [{\citenamefont {Chen}\ \emph {et~al.}(2011)\citenamefont {Chen},
  \citenamefont {Gu},\ and\ \citenamefont {Wen}}]{Chen2011a}%
  \BibitemOpen
  \bibfield  {author} {\bibinfo {author} {\bibfnamefont {X.}~\bibnamefont
  {Chen}}, \bibinfo {author} {\bibfnamefont {Z.-C.}\ \bibnamefont {Gu}}, \ and\
  \bibinfo {author} {\bibfnamefont {X.-G.}\ \bibnamefont {Wen}},\ }\href
  {\doibase 10.1103/PhysRevB.84.235128} {\bibfield  {journal} {\bibinfo
  {journal} {Phys. Rev. B}\ }\textbf {\bibinfo {volume} {84}},\ \bibinfo
  {pages} {235128} (\bibinfo {year} {2011})}\BibitemShut {NoStop}%
\bibitem [{\citenamefont {Schuch}\ \emph {et~al.}(2011)\citenamefont {Schuch},
  \citenamefont {P\'erez-Garc\'~ia},\ and\ \citenamefont {Cirac}}]{Schuch2011}%
  \BibitemOpen
  \bibfield  {author} {\bibinfo {author} {\bibfnamefont {N.}~\bibnamefont
  {Schuch}}, \bibinfo {author} {\bibfnamefont {D.}~\bibnamefont
  {P\'erez-Garc\'~ia}}, \ and\ \bibinfo {author} {\bibfnamefont
  {I.}~\bibnamefont {Cirac}},\ }\href {\doibase 10.1103/PhysRevB.84.165139}
  {\bibfield  {journal} {\bibinfo  {journal} {Phys. Rev. B}\ }\textbf {\bibinfo
  {volume} {84}},\ \bibinfo {pages} {165139} (\bibinfo {year}
  {2011})}\BibitemShut {NoStop}%
\bibitem [{\citenamefont {Buyers}\ \emph {et~al.}(1986)\citenamefont {Buyers},
  \citenamefont {Morra}, \citenamefont {Armstrong}, \citenamefont {Hogan},
  \citenamefont {Gerlach},\ and\ \citenamefont {Hirakawa}}]{Buyers1986}%
  \BibitemOpen
  \bibfield  {author} {\bibinfo {author} {\bibfnamefont {W.~J.~L.}\
  \bibnamefont {Buyers}}, \bibinfo {author} {\bibfnamefont {R.~M.}\
  \bibnamefont {Morra}}, \bibinfo {author} {\bibfnamefont {R.~L.}\ \bibnamefont
  {Armstrong}}, \bibinfo {author} {\bibfnamefont {M.~J.}\ \bibnamefont
  {Hogan}}, \bibinfo {author} {\bibfnamefont {P.}~\bibnamefont {Gerlach}}, \
  and\ \bibinfo {author} {\bibfnamefont {K.}~\bibnamefont {Hirakawa}},\ }\href
  {\doibase 10.1103/PhysRevLett.56.371} {\bibfield  {journal} {\bibinfo
  {journal} {Phys. Rev. Lett.}\ }\textbf {\bibinfo {volume} {56}},\ \bibinfo
  {pages} {371} (\bibinfo {year} {1986})}\BibitemShut {NoStop}%
\bibitem [{\citenamefont {Khanikaev}\ \emph {et~al.}(2013)\citenamefont
  {Khanikaev}, \citenamefont {Hossein~Mousavi}, \citenamefont {Tse},
  \citenamefont {Kargarian}, \citenamefont {MacDonald},\ and\ \citenamefont
  {Shvets}}]{Gennady2013}%
  \BibitemOpen
  \bibfield  {author} {\bibinfo {author} {\bibfnamefont {A.~B.}\ \bibnamefont
  {Khanikaev}}, \bibinfo {author} {\bibfnamefont {S.}~\bibnamefont
  {Hossein~Mousavi}}, \bibinfo {author} {\bibfnamefont {W.-K.}\ \bibnamefont
  {Tse}}, \bibinfo {author} {\bibfnamefont {M.}~\bibnamefont {Kargarian}},
  \bibinfo {author} {\bibfnamefont {A.~H.}\ \bibnamefont {MacDonald}}, \ and\
  \bibinfo {author} {\bibfnamefont {G.}~\bibnamefont {Shvets}},\ }\href
  {http://dx.doi.org/10.1038/nmat3520} {\bibfield  {journal} {\bibinfo
  {journal} {Nat Mater}\ }\textbf {\bibinfo {volume} {12}},\ \bibinfo {pages}
  {233} (\bibinfo {year} {2013})}\BibitemShut {NoStop}%
\bibitem [{\citenamefont {Kane}\ and\ \citenamefont
  {Lubensky}(2014)}]{Kane2014}%
  \BibitemOpen
  \bibfield  {author} {\bibinfo {author} {\bibfnamefont {C.~L.}\ \bibnamefont
  {Kane}}\ and\ \bibinfo {author} {\bibfnamefont {T.~C.}\ \bibnamefont
  {Lubensky}},\ }\href {http://dx.doi.org/10.1038/nphys2835} {\bibfield
  {journal} {\bibinfo  {journal} {Nat Phys}\ }\textbf {\bibinfo {volume}
  {10}},\ \bibinfo {pages} {39} (\bibinfo {year} {2014})}\BibitemShut {NoStop}%
\bibitem [{\citenamefont {Rocklin}\ \emph {et~al.}(2016)\citenamefont
  {Rocklin}, \citenamefont {Chen}, \citenamefont {Falk}, \citenamefont
  {Vitelli},\ and\ \citenamefont {Lubensky}}]{PhysRevLett.116.135503}%
  \BibitemOpen
  \bibfield  {author} {\bibinfo {author} {\bibfnamefont {D.~Z.}\ \bibnamefont
  {Rocklin}}, \bibinfo {author} {\bibfnamefont {B.~G.}\ \bibnamefont {Chen}},
  \bibinfo {author} {\bibfnamefont {M.}~\bibnamefont {Falk}}, \bibinfo {author}
  {\bibfnamefont {V.}~\bibnamefont {Vitelli}}, \ and\ \bibinfo {author}
  {\bibfnamefont {T.~C.}\ \bibnamefont {Lubensky}},\ }\href {\doibase
  10.1103/PhysRevLett.116.135503} {\bibfield  {journal} {\bibinfo  {journal}
  {Phys. Rev. Lett.}\ }\textbf {\bibinfo {volume} {116}},\ \bibinfo {pages}
  {135503} (\bibinfo {year} {2016})}\BibitemShut {NoStop}%
\bibitem [{\citenamefont {S{\"u}sstrunk}\ and\ \citenamefont
  {Huber}(2015)}]{Huber2015}%
  \BibitemOpen
  \bibfield  {author} {\bibinfo {author} {\bibfnamefont {R.}~\bibnamefont
  {S{\"u}sstrunk}}\ and\ \bibinfo {author} {\bibfnamefont {S.~D.}\ \bibnamefont
  {Huber}},\ }\href
  {http://science.sciencemag.org/content/349/6243/47.abstract} {\bibfield
  {journal} {\bibinfo  {journal} {Science}\ }\textbf {\bibinfo {volume}
  {349}},\ \bibinfo {pages} {47} (\bibinfo {year} {2015})}\BibitemShut
  {NoStop}%
\bibitem [{\citenamefont {Skirlo}\ \emph {et~al.}(2015)\citenamefont {Skirlo},
  \citenamefont {Lu}, \citenamefont {Igarashi}, \citenamefont {Yan},
  \citenamefont {Joannopoulos},\ and\ \citenamefont {Solja\ifmmode
  \check{c}\else \v{c}\fi{}i\ifmmode~\acute{c}\else
  \'{c}\fi{}}}]{PhotonicsExp1}%
  \BibitemOpen
  \bibfield  {author} {\bibinfo {author} {\bibfnamefont {S.~A.}\ \bibnamefont
  {Skirlo}}, \bibinfo {author} {\bibfnamefont {L.}~\bibnamefont {Lu}}, \bibinfo
  {author} {\bibfnamefont {Y.}~\bibnamefont {Igarashi}}, \bibinfo {author}
  {\bibfnamefont {Q.}~\bibnamefont {Yan}}, \bibinfo {author} {\bibfnamefont
  {J.}~\bibnamefont {Joannopoulos}}, \ and\ \bibinfo {author} {\bibfnamefont
  {M.}~\bibnamefont {Solja\ifmmode \check{c}\else
  \v{c}\fi{}i\ifmmode~\acute{c}\else \'{c}\fi{}}},\ }\href {\doibase
  10.1103/PhysRevLett.115.253901} {\bibfield  {journal} {\bibinfo  {journal}
  {Phys. Rev. Lett.}\ }\textbf {\bibinfo {volume} {115}},\ \bibinfo {pages}
  {253901} (\bibinfo {year} {2015})}\BibitemShut {NoStop}%
\bibitem [{\citenamefont {Rechtsman}\ \emph {et~al.}(2013)\citenamefont
  {Rechtsman}, \citenamefont {Zeuner}, \citenamefont {Plotnik}, \citenamefont
  {Lumer}, \citenamefont {Podolsky}, \citenamefont {Dreisow}, \citenamefont
  {Nolte}, \citenamefont {Segev},\ and\ \citenamefont
  {Szameit}}]{PhotonicsExp2}%
  \BibitemOpen
  \bibfield  {author} {\bibinfo {author} {\bibfnamefont {M.~C.}\ \bibnamefont
  {Rechtsman}}, \bibinfo {author} {\bibfnamefont {J.~M.}\ \bibnamefont
  {Zeuner}}, \bibinfo {author} {\bibfnamefont {Y.}~\bibnamefont {Plotnik}},
  \bibinfo {author} {\bibfnamefont {Y.}~\bibnamefont {Lumer}}, \bibinfo
  {author} {\bibfnamefont {D.}~\bibnamefont {Podolsky}}, \bibinfo {author}
  {\bibfnamefont {F.}~\bibnamefont {Dreisow}}, \bibinfo {author} {\bibfnamefont
  {S.}~\bibnamefont {Nolte}}, \bibinfo {author} {\bibfnamefont
  {M.}~\bibnamefont {Segev}}, \ and\ \bibinfo {author} {\bibfnamefont
  {A.}~\bibnamefont {Szameit}},\ }\href {http://dx.doi.org/10.1038/nature12066}
  {\bibfield  {journal} {\bibinfo  {journal} {Nature}\ }\textbf {\bibinfo
  {volume} {496}},\ \bibinfo {pages} {196} (\bibinfo {year}
  {2013})}\BibitemShut {NoStop}%
\bibitem [{\citenamefont {Kraus}\ \emph {et~al.}(2012)\citenamefont {Kraus},
  \citenamefont {Lahini}, \citenamefont {Ringel}, \citenamefont {Verbin},\ and\
  \citenamefont {Zilberberg}}]{PhotonicsExp3}%
  \BibitemOpen
  \bibfield  {author} {\bibinfo {author} {\bibfnamefont {Y.~E.}\ \bibnamefont
  {Kraus}}, \bibinfo {author} {\bibfnamefont {Y.}~\bibnamefont {Lahini}},
  \bibinfo {author} {\bibfnamefont {Z.}~\bibnamefont {Ringel}}, \bibinfo
  {author} {\bibfnamefont {M.}~\bibnamefont {Verbin}}, \ and\ \bibinfo {author}
  {\bibfnamefont {O.}~\bibnamefont {Zilberberg}},\ }\href {\doibase
  10.1103/PhysRevLett.109.106402} {\bibfield  {journal} {\bibinfo  {journal}
  {Phys. Rev. Lett.}\ }\textbf {\bibinfo {volume} {109}},\ \bibinfo {pages}
  {106402} (\bibinfo {year} {2012})}\BibitemShut {NoStop}%
\bibitem [{\citenamefont {Paulose}\ \emph {et~al.}(2015)\citenamefont
  {Paulose}, \citenamefont {Meeussen},\ and\ \citenamefont
  {Vitelli}}]{Paulose23062015}%
  \BibitemOpen
  \bibfield  {author} {\bibinfo {author} {\bibfnamefont {J.}~\bibnamefont
  {Paulose}}, \bibinfo {author} {\bibfnamefont {A.~S.}\ \bibnamefont
  {Meeussen}}, \ and\ \bibinfo {author} {\bibfnamefont {V.}~\bibnamefont
  {Vitelli}},\ }\href {\doibase 10.1073/pnas.1502939112} {\bibfield  {journal}
  {\bibinfo  {journal} {Proceedings of the National Academy of Sciences}\
  }\textbf {\bibinfo {volume} {112}},\ \bibinfo {pages} {7639} (\bibinfo {year}
  {2015})},\ \Eprint
  {http://arxiv.org/abs/http://www.pnas.org/content/112/25/7639.full.pdf}
  {http://www.pnas.org/content/112/25/7639.full.pdf} \BibitemShut {NoStop}%
\bibitem [{\citenamefont {Chen}\ \emph
  {et~al.}(2014{\natexlab{a}})\citenamefont {Chen}, \citenamefont {Upadhyaya},\
  and\ \citenamefont {Vitelli}}]{Chen09092014}%
  \BibitemOpen
  \bibfield  {author} {\bibinfo {author} {\bibfnamefont {B.~G.-g.}\
  \bibnamefont {Chen}}, \bibinfo {author} {\bibfnamefont {N.}~\bibnamefont
  {Upadhyaya}}, \ and\ \bibinfo {author} {\bibfnamefont {V.}~\bibnamefont
  {Vitelli}},\ }\href {\doibase 10.1073/pnas.1405969111} {\bibfield  {journal}
  {\bibinfo  {journal} {Proceedings of the National Academy of Sciences}\
  }\textbf {\bibinfo {volume} {111}},\ \bibinfo {pages} {13004} (\bibinfo
  {year} {2014}{\natexlab{a}})},\ \Eprint
  {http://arxiv.org/abs/http://www.pnas.org/content/111/36/13004.full.pdf}
  {http://www.pnas.org/content/111/36/13004.full.pdf} \BibitemShut {NoStop}%
\bibitem [{\citenamefont {{Hafezi}}\ \emph {et~al.}(2011)\citenamefont
  {{Hafezi}}, \citenamefont {{Demler}}, \citenamefont {{Lukin}},\ and\
  \citenamefont {{Taylor}}}]{Hafezi2011}%
  \BibitemOpen
  \bibfield  {author} {\bibinfo {author} {\bibfnamefont {M.}~\bibnamefont
  {{Hafezi}}}, \bibinfo {author} {\bibfnamefont {E.~A.}\ \bibnamefont
  {{Demler}}}, \bibinfo {author} {\bibfnamefont {M.~D.}\ \bibnamefont
  {{Lukin}}}, \ and\ \bibinfo {author} {\bibfnamefont {J.~M.}\ \bibnamefont
  {{Taylor}}},\ }\href {\doibase 10.1038/nphys2063} {\bibfield  {journal}
  {\bibinfo  {journal} {Nature Physics}\ }\textbf {\bibinfo {volume} {7}},\
  \bibinfo {pages} {907} (\bibinfo {year} {2011})},\ \Eprint
  {http://arxiv.org/abs/1102.3256} {arXiv:1102.3256 [quant-ph]} \BibitemShut
  {NoStop}%
\bibitem [{\citenamefont {Affleck}\ \emph {et~al.}(1988)\citenamefont
  {Affleck}, \citenamefont {Kennedy}, \citenamefont {Lieb},\ and\ \citenamefont
  {Tasaki}}]{AKLT1988}%
  \BibitemOpen
  \bibfield  {author} {\bibinfo {author} {\bibfnamefont {I.}~\bibnamefont
  {Affleck}}, \bibinfo {author} {\bibfnamefont {T.}~\bibnamefont {Kennedy}},
  \bibinfo {author} {\bibfnamefont {E.~H.}\ \bibnamefont {Lieb}}, \ and\
  \bibinfo {author} {\bibfnamefont {H.}~\bibnamefont {Tasaki}},\ }\href
  {http://projecteuclid.org/euclid.cmp/1104161001} {\bibfield  {journal}
  {\bibinfo  {journal} {Comm. Math. Phys.}\ }\textbf {\bibinfo {volume}
  {115}},\ \bibinfo {pages} {477} (\bibinfo {year} {1988})}\BibitemShut
  {NoStop}%
\bibitem [{\citenamefont {Pollmann}\ \emph {et~al.}(2012)\citenamefont
  {Pollmann}, \citenamefont {Berg}, \citenamefont {Turner},\ and\ \citenamefont
  {Oshikawa}}]{Pollmann2012}%
  \BibitemOpen
  \bibfield  {author} {\bibinfo {author} {\bibfnamefont {F.}~\bibnamefont
  {Pollmann}}, \bibinfo {author} {\bibfnamefont {E.}~\bibnamefont {Berg}},
  \bibinfo {author} {\bibfnamefont {A.~M.}\ \bibnamefont {Turner}}, \ and\
  \bibinfo {author} {\bibfnamefont {M.}~\bibnamefont {Oshikawa}},\ }\href
  {\doibase 10.1103/PhysRevB.85.075125} {\bibfield  {journal} {\bibinfo
  {journal} {Phys. Rev. B}\ }\textbf {\bibinfo {volume} {85}},\ \bibinfo
  {pages} {075125} (\bibinfo {year} {2012})}\BibitemShut {NoStop}%
\bibitem [{\citenamefont {Geraedts}\ and\ \citenamefont
  {Motrunich}(2013)}]{Geraedts2013}%
  \BibitemOpen
  \bibfield  {author} {\bibinfo {author} {\bibfnamefont {S.~D.}\ \bibnamefont
  {Geraedts}}\ and\ \bibinfo {author} {\bibfnamefont {O.~I.}\ \bibnamefont
  {Motrunich}},\ }\href {\doibase http://dx.doi.org/10.1016/j.aop.2013.03.017}
  {\bibfield  {journal} {\bibinfo  {journal} {Annals of Physics}\ }\textbf
  {\bibinfo {volume} {334}},\ \bibinfo {pages} {288 } (\bibinfo {year}
  {2013})}\BibitemShut {NoStop}%
\bibitem [{\citenamefont {van Hove}(1950)}]{VanHove1950}%
  \BibitemOpen
  \bibfield  {author} {\bibinfo {author} {\bibfnamefont {L.}~\bibnamefont {van
  Hove}},\ }\href {\doibase http://dx.doi.org/10.1016/0031-8914(50)90072-3}
  {\bibfield  {journal} {\bibinfo  {journal} {Physica}\ }\textbf {\bibinfo
  {volume} {16}},\ \bibinfo {pages} {137 } (\bibinfo {year}
  {1950})}\BibitemShut {NoStop}%
\bibitem [{\citenamefont {Itzykson}\ and\ \citenamefont
  {Drouffe}(1991)}]{Cohomology1991}%
  \BibitemOpen
  \bibfield  {author} {\bibinfo {author} {\bibfnamefont {C.}~\bibnamefont
  {Itzykson}}\ and\ \bibinfo {author} {\bibfnamefont {J.}~\bibnamefont
  {Drouffe}},\ }\href {https://books.google.co.uk/books?id=cKdTrJPBAyUC} {\emph
  {\bibinfo {title} {Statistical Field Theory: Volume 1, From Brownian Motion
  to Renormalization and Lattice Gauge Theory}}}\ (\bibinfo  {publisher}
  {Cambridge University Press},\ \bibinfo {year} {1991})\BibitemShut {NoStop}%
\bibitem [{Note1()}]{Note1}%
  \BibitemOpen
  \bibinfo {note} {We remark that one can consider more general couplings, such
  as those of the Ashkin--Teller model \cite {Baxter}, as long as the set of
  order parameters in a phase is unchanged. Our choice of two decoupled Ising
  models is made for pedagogical purposes.}\BibitemShut {Stop}%
\bibitem [{\citenamefont {Savit}(1980)}]{Savit}%
  \BibitemOpen
  \bibfield  {author} {\bibinfo {author} {\bibfnamefont {R.}~\bibnamefont
  {Savit}},\ }\href {\doibase 10.1103/RevModPhys.52.453} {\bibfield  {journal}
  {\bibinfo  {journal} {Rev. Mod. Phys.}\ }\textbf {\bibinfo {volume} {52}},\
  \bibinfo {pages} {453} (\bibinfo {year} {1980})}\BibitemShut {NoStop}%
\bibitem [{\citenamefont {Chen}\ \emph
  {et~al.}(2014{\natexlab{b}})\citenamefont {Chen}, \citenamefont {Lu},\ and\
  \citenamefont {Vishwanath}}]{chen2014symmetry}%
  \BibitemOpen
  \bibfield  {author} {\bibinfo {author} {\bibfnamefont {X.}~\bibnamefont
  {Chen}}, \bibinfo {author} {\bibfnamefont {Y.-M.}\ \bibnamefont {Lu}}, \ and\
  \bibinfo {author} {\bibfnamefont {A.}~\bibnamefont {Vishwanath}},\
  }\href@noop {} {\bibfield  {journal} {\bibinfo  {journal} {Nature
  communications}\ }\textbf {\bibinfo {volume} {5}} (\bibinfo {year}
  {2014}{\natexlab{b}})}\BibitemShut {NoStop}%
\bibitem [{\citenamefont {Kogut}(1979)}]{Kogut}%
  \BibitemOpen
  \bibfield  {author} {\bibinfo {author} {\bibfnamefont {J.~B.}\ \bibnamefont
  {Kogut}},\ }\href {\doibase 10.1103/RevModPhys.51.659} {\bibfield  {journal}
  {\bibinfo  {journal} {Rev. Mod. Phys.}\ }\textbf {\bibinfo {volume} {51}},\
  \bibinfo {pages} {659} (\bibinfo {year} {1979})}\BibitemShut {NoStop}%
\bibitem [{\citenamefont {Ringel}\ and\ \citenamefont
  {Simon}(2015)}]{Ringel2015}%
  \BibitemOpen
  \bibfield  {author} {\bibinfo {author} {\bibfnamefont {Z.}~\bibnamefont
  {Ringel}}\ and\ \bibinfo {author} {\bibfnamefont {S.~H.}\ \bibnamefont
  {Simon}},\ }\href {\doibase 10.1103/PhysRevB.91.195117} {\bibfield  {journal}
  {\bibinfo  {journal} {Phys. Rev. B}\ }\textbf {\bibinfo {volume} {91}},\
  \bibinfo {pages} {195117} (\bibinfo {year} {2015})}\BibitemShut {NoStop}%
\bibitem [{\citenamefont {Scaffidi}\ and\ \citenamefont
  {Ringel}(2016)}]{Scaffidi2016}%
  \BibitemOpen
  \bibfield  {author} {\bibinfo {author} {\bibfnamefont {T.}~\bibnamefont
  {Scaffidi}}\ and\ \bibinfo {author} {\bibfnamefont {Z.}~\bibnamefont
  {Ringel}},\ }\href {\doibase 10.1103/PhysRevB.93.115105} {\bibfield
  {journal} {\bibinfo  {journal} {Phys. Rev. B}\ }\textbf {\bibinfo {volume}
  {93}},\ \bibinfo {pages} {115105} (\bibinfo {year} {2016})}\BibitemShut
  {NoStop}%
\bibitem [{\citenamefont {Fradkin}\ and\ \citenamefont
  {Shenker}(1979)}]{Fradkin1979}%
  \BibitemOpen
  \bibfield  {author} {\bibinfo {author} {\bibfnamefont {E.}~\bibnamefont
  {Fradkin}}\ and\ \bibinfo {author} {\bibfnamefont {S.~H.}\ \bibnamefont
  {Shenker}},\ }\href {\doibase 10.1103/PhysRevD.19.3682} {\bibfield  {journal}
  {\bibinfo  {journal} {Phys. Rev. D}\ }\textbf {\bibinfo {volume} {19}},\
  \bibinfo {pages} {3682} (\bibinfo {year} {1979})}\BibitemShut {NoStop}%
\bibitem [{\citenamefont {Baxter}(2007)}]{Baxter}%
  \BibitemOpen
  \bibfield  {author} {\bibinfo {author} {\bibfnamefont {R.}~\bibnamefont
  {Baxter}},\ }\href {https://books.google.co.uk/books?id=G3owDULfBuEC} {\emph
  {\bibinfo {title} {Exactly Solved Models in Statistical Mechanics}}},\ Dover
  books on physics\ (\bibinfo  {publisher} {Dover Publications},\ \bibinfo
  {year} {2007})\BibitemShut {NoStop}%
\bibitem [{\citenamefont {Nienhuis}(1984)}]{Nienhuis1984}%
  \BibitemOpen
  \bibfield  {author} {\bibinfo {author} {\bibfnamefont {B.}~\bibnamefont
  {Nienhuis}},\ }\href {\doibase 10.1007/BF01009437} {\bibfield  {journal}
  {\bibinfo  {journal} {Journal of Statistical Physics}\ }\textbf {\bibinfo
  {volume} {34}},\ \bibinfo {pages} {731} (\bibinfo {year} {1984})}\BibitemShut
  {NoStop}%
\bibitem [{\citenamefont {Lu}\ and\ \citenamefont
  {Vishwanath}(2012)}]{YuanMing2012}%
  \BibitemOpen
  \bibfield  {author} {\bibinfo {author} {\bibfnamefont {Y.-M.}\ \bibnamefont
  {Lu}}\ and\ \bibinfo {author} {\bibfnamefont {A.}~\bibnamefont
  {Vishwanath}},\ }\href {\doibase 10.1103/PhysRevB.86.125119} {\bibfield
  {journal} {\bibinfo  {journal} {Phys. Rev. B}\ }\textbf {\bibinfo {volume}
  {86}},\ \bibinfo {pages} {125119} (\bibinfo {year} {2012})}\BibitemShut
  {NoStop}%
\bibitem [{\citenamefont {Else}\ and\ \citenamefont {Nayak}(2014)}]{Nayak2014}%
  \BibitemOpen
  \bibfield  {author} {\bibinfo {author} {\bibfnamefont {D.~V.}\ \bibnamefont
  {Else}}\ and\ \bibinfo {author} {\bibfnamefont {C.}~\bibnamefont {Nayak}},\
  }\href {\doibase 10.1103/PhysRevB.90.235137} {\bibfield  {journal} {\bibinfo
  {journal} {Phys. Rev. B}\ }\textbf {\bibinfo {volume} {90}},\ \bibinfo
  {pages} {235137} (\bibinfo {year} {2014})}\BibitemShut {NoStop}%
\bibitem [{\citenamefont {{Reshetikhin}}(2010)}]{Reshetikhin}%
  \BibitemOpen
  \bibfield  {author} {\bibinfo {author} {\bibfnamefont {N.}~\bibnamefont
  {{Reshetikhin}}},\ }\href@noop {} {\bibfield  {journal} {\bibinfo  {journal}
  {ArXiv e-prints}\ } (\bibinfo {year} {2010})},\ \Eprint
  {http://arxiv.org/abs/1010.5031} {arXiv:1010.5031 [math-ph]} \BibitemShut
  {NoStop}%
\bibitem [{\citenamefont {Noh}\ and\ \citenamefont {Kim}(1996)}]{Kim}%
  \BibitemOpen
  \bibfield  {author} {\bibinfo {author} {\bibfnamefont {J.~D.}\ \bibnamefont
  {Noh}}\ and\ \bibinfo {author} {\bibfnamefont {D.}~\bibnamefont {Kim}},\
  }\href {\doibase 10.1103/PhysRevE.53.3225} {\bibfield  {journal} {\bibinfo
  {journal} {Phys. Rev. E}\ }\textbf {\bibinfo {volume} {53}},\ \bibinfo
  {pages} {3225} (\bibinfo {year} {1996})}\BibitemShut {NoStop}%
\bibitem [{\citenamefont {Balian}\ \emph {et~al.}(1975)\citenamefont {Balian},
  \citenamefont {Drouffe},\ and\ \citenamefont {Itzykson}}]{Balian}%
  \BibitemOpen
  \bibfield  {author} {\bibinfo {author} {\bibfnamefont {R.}~\bibnamefont
  {Balian}}, \bibinfo {author} {\bibfnamefont {J.~M.}\ \bibnamefont {Drouffe}},
  \ and\ \bibinfo {author} {\bibfnamefont {C.}~\bibnamefont {Itzykson}},\
  }\href {\doibase 10.1103/PhysRevD.11.2098} {\bibfield  {journal} {\bibinfo
  {journal} {Phys. Rev. D}\ }\textbf {\bibinfo {volume} {11}},\ \bibinfo
  {pages} {2098} (\bibinfo {year} {1975})}\BibitemShut {NoStop}%
\bibitem [{\citenamefont {Wang}\ \emph {et~al.}(2015)\citenamefont {Wang},
  \citenamefont {Santos},\ and\ \citenamefont {Wen}}]{Juven2015}%
  \BibitemOpen
  \bibfield  {author} {\bibinfo {author} {\bibfnamefont {J.~C.}\ \bibnamefont
  {Wang}}, \bibinfo {author} {\bibfnamefont {L.~H.}\ \bibnamefont {Santos}}, \
  and\ \bibinfo {author} {\bibfnamefont {X.-G.}\ \bibnamefont {Wen}},\ }\href
  {\doibase 10.1103/PhysRevB.91.195134} {\bibfield  {journal} {\bibinfo
  {journal} {Phys. Rev. B}\ }\textbf {\bibinfo {volume} {91}},\ \bibinfo
  {pages} {195134} (\bibinfo {year} {2015})}\BibitemShut {NoStop}%
\bibitem [{\citenamefont {Coste}\ \emph {et~al.}(2000)\citenamefont {Coste},
  \citenamefont {Gannon},\ and\ \citenamefont {Ruelle}}]{ModularData}%
  \BibitemOpen
  \bibfield  {author} {\bibinfo {author} {\bibfnamefont {A.}~\bibnamefont
  {Coste}}, \bibinfo {author} {\bibfnamefont {T.}~\bibnamefont {Gannon}}, \
  and\ \bibinfo {author} {\bibfnamefont {P.}~\bibnamefont {Ruelle}},\ }\href
  {\doibase http://dx.doi.org/10.1016/S0550-3213(00)00285-6} {\bibfield
  {journal} {\bibinfo  {journal} {Nuclear Physics B}\ }\textbf {\bibinfo
  {volume} {581}},\ \bibinfo {pages} {679 } (\bibinfo {year}
  {2000})}\BibitemShut {NoStop}%
\bibitem [{\citenamefont {Laughlin}(1981)}]{Laughlin1981}%
  \BibitemOpen
  \bibfield  {author} {\bibinfo {author} {\bibfnamefont {R.~B.}\ \bibnamefont
  {Laughlin}},\ }\href {\doibase 10.1103/PhysRevB.23.5632} {\bibfield
  {journal} {\bibinfo  {journal} {Phys. Rev. B}\ }\textbf {\bibinfo {volume}
  {23}},\ \bibinfo {pages} {5632} (\bibinfo {year} {1981})}\BibitemShut
  {NoStop}%
\bibitem [{\citenamefont {Levin}\ and\ \citenamefont {Gu}(2012)}]{Levin2012}%
  \BibitemOpen
  \bibfield  {author} {\bibinfo {author} {\bibfnamefont {M.}~\bibnamefont
  {Levin}}\ and\ \bibinfo {author} {\bibfnamefont {Z.-C.}\ \bibnamefont {Gu}},\
  }\href {\doibase 10.1103/PhysRevB.86.115109} {\bibfield  {journal} {\bibinfo
  {journal} {Phys. Rev. B}\ }\textbf {\bibinfo {volume} {86}},\ \bibinfo
  {pages} {115109} (\bibinfo {year} {2012})}\BibitemShut {NoStop}%
\bibitem [{\citenamefont {Hatcher}(2002)}]{Hatcher2002}%
  \BibitemOpen
  \bibfield  {author} {\bibinfo {author} {\bibfnamefont {A.}~\bibnamefont
  {Hatcher}},\ }\href {http://books.google.it/books?id=BjKs86kosqgC} {\emph
  {\bibinfo {title} {Algebraic Topology}}}\ (\bibinfo  {publisher} {Cambridge
  University Press},\ \bibinfo {year} {2002})\BibitemShut {NoStop}%
\bibitem [{\citenamefont {{Kapustin}}(2014)}]{Kapustin2014}%
  \BibitemOpen
  \bibfield  {author} {\bibinfo {author} {\bibfnamefont {A.}~\bibnamefont
  {{Kapustin}}},\ }\href@noop {} {\bibfield  {journal} {\bibinfo  {journal}
  {ArXiv e-prints}\ } (\bibinfo {year} {2014})},\ \Eprint
  {http://arxiv.org/abs/1403.1467} {arXiv:1403.1467 [cond-mat.str-el]}
  \BibitemShut {NoStop}%
\bibitem [{\citenamefont {Yung}\ and\ \citenamefont
  {Batchelor}(1995)}]{YUNG1995}%
  \BibitemOpen
  \bibfield  {author} {\bibinfo {author} {\bibfnamefont {C.}~\bibnamefont
  {Yung}}\ and\ \bibinfo {author} {\bibfnamefont {M.}~\bibnamefont
  {Batchelor}},\ }\href {\doibase
  http://dx.doi.org/10.1016/0550-3213(94)00448-N} {\bibfield  {journal}
  {\bibinfo  {journal} {Nuclear Physics B}\ }\textbf {\bibinfo {volume}
  {435}},\ \bibinfo {pages} {430 } (\bibinfo {year} {1995})}\BibitemShut
  {NoStop}%
\bibitem [{\citenamefont {Honda}\ and\ \citenamefont
  {Horiguchi}(1997)}]{Honda1997}%
  \BibitemOpen
  \bibfield  {author} {\bibinfo {author} {\bibfnamefont {Y.}~\bibnamefont
  {Honda}}\ and\ \bibinfo {author} {\bibfnamefont {T.}~\bibnamefont
  {Horiguchi}},\ }\href {\doibase 10.1103/PhysRevE.56.3920} {\bibfield
  {journal} {\bibinfo  {journal} {Phys. Rev. E}\ }\textbf {\bibinfo {volume}
  {56}},\ \bibinfo {pages} {3920} (\bibinfo {year} {1997})}\BibitemShut
  {NoStop}%
\bibitem [{\citenamefont {You}\ \emph {et~al.}(2016)\citenamefont {You},
  \citenamefont {Bi}, \citenamefont {Mao},\ and\ \citenamefont {Xu}}]{You2016}%
  \BibitemOpen
  \bibfield  {author} {\bibinfo {author} {\bibfnamefont {Y.-Z.}\ \bibnamefont
  {You}}, \bibinfo {author} {\bibfnamefont {Z.}~\bibnamefont {Bi}}, \bibinfo
  {author} {\bibfnamefont {D.}~\bibnamefont {Mao}}, \ and\ \bibinfo {author}
  {\bibfnamefont {C.}~\bibnamefont {Xu}},\ }\href {\doibase
  10.1103/PhysRevB.93.125101} {\bibfield  {journal} {\bibinfo  {journal} {Phys.
  Rev. B}\ }\textbf {\bibinfo {volume} {93}},\ \bibinfo {pages} {125101}
  (\bibinfo {year} {2016})}\BibitemShut {NoStop}%
\bibitem [{\citenamefont {Wang}\ \emph {et~al.}(2006)\citenamefont {Wang},
  \citenamefont {Nisoli}, \citenamefont {Freitas}, \citenamefont {Li},
  \citenamefont {McConville}, \citenamefont {Cooley}, \citenamefont {Lund},
  \citenamefont {Samarth}, \citenamefont {Leighton}, \citenamefont {Crespi},\
  and\ \citenamefont {Schiffer}}]{Wang2006}%
  \BibitemOpen
  \bibfield  {author} {\bibinfo {author} {\bibfnamefont {R.~F.}\ \bibnamefont
  {Wang}}, \bibinfo {author} {\bibfnamefont {C.}~\bibnamefont {Nisoli}},
  \bibinfo {author} {\bibfnamefont {R.~S.}\ \bibnamefont {Freitas}}, \bibinfo
  {author} {\bibfnamefont {J.}~\bibnamefont {Li}}, \bibinfo {author}
  {\bibfnamefont {W.}~\bibnamefont {McConville}}, \bibinfo {author}
  {\bibfnamefont {B.~J.}\ \bibnamefont {Cooley}}, \bibinfo {author}
  {\bibfnamefont {M.~S.}\ \bibnamefont {Lund}}, \bibinfo {author}
  {\bibfnamefont {N.}~\bibnamefont {Samarth}}, \bibinfo {author} {\bibfnamefont
  {C.}~\bibnamefont {Leighton}}, \bibinfo {author} {\bibfnamefont {V.~H.}\
  \bibnamefont {Crespi}}, \ and\ \bibinfo {author} {\bibfnamefont
  {P.}~\bibnamefont {Schiffer}},\ }\href
  {http://dx.doi.org/10.1038/nature04447} {\bibfield  {journal} {\bibinfo
  {journal} {Nature}\ }\textbf {\bibinfo {volume} {439}},\ \bibinfo {pages}
  {303} (\bibinfo {year} {2006})}\BibitemShut {NoStop}%
\bibitem [{\citenamefont {Qi}\ \emph {et~al.}(2008)\citenamefont {Qi},
  \citenamefont {Brintlinger},\ and\ \citenamefont {Cumings}}]{Cumings2008}%
  \BibitemOpen
  \bibfield  {author} {\bibinfo {author} {\bibfnamefont {Y.}~\bibnamefont
  {Qi}}, \bibinfo {author} {\bibfnamefont {T.}~\bibnamefont {Brintlinger}}, \
  and\ \bibinfo {author} {\bibfnamefont {J.}~\bibnamefont {Cumings}},\ }\href
  {\doibase 10.1103/PhysRevB.77.094418} {\bibfield  {journal} {\bibinfo
  {journal} {Phys. Rev. B}\ }\textbf {\bibinfo {volume} {77}},\ \bibinfo
  {pages} {094418} (\bibinfo {year} {2008})}\BibitemShut {NoStop}%
\bibitem [{\citenamefont {Nisoli}\ \emph {et~al.}(2013)\citenamefont {Nisoli},
  \citenamefont {Moessner},\ and\ \citenamefont {Schiffer}}]{Roderich2013}%
  \BibitemOpen
  \bibfield  {author} {\bibinfo {author} {\bibfnamefont {C.}~\bibnamefont
  {Nisoli}}, \bibinfo {author} {\bibfnamefont {R.}~\bibnamefont {Moessner}}, \
  and\ \bibinfo {author} {\bibfnamefont {P.}~\bibnamefont {Schiffer}},\ }\href
  {\doibase 10.1103/RevModPhys.85.1473} {\bibfield  {journal} {\bibinfo
  {journal} {Rev. Mod. Phys.}\ }\textbf {\bibinfo {volume} {85}},\ \bibinfo
  {pages} {1473} (\bibinfo {year} {2013})}\BibitemShut {NoStop}%
\bibitem [{\citenamefont {Blunt}\ \emph {et~al.}(2008)\citenamefont {Blunt},
  \citenamefont {Russell}, \citenamefont {Gim{\'e}nez-L{\'o}pez}, \citenamefont
  {Garrahan}, \citenamefont {Lin}, \citenamefont {Schr{\"o}der}, \citenamefont
  {Champness},\ and\ \citenamefont {Beton}}]{Blunt}%
  \BibitemOpen
  \bibfield  {author} {\bibinfo {author} {\bibfnamefont {M.~O.}\ \bibnamefont
  {Blunt}}, \bibinfo {author} {\bibfnamefont {J.~C.}\ \bibnamefont {Russell}},
  \bibinfo {author} {\bibfnamefont {M.~d.~C.}\ \bibnamefont
  {Gim{\'e}nez-L{\'o}pez}}, \bibinfo {author} {\bibfnamefont {J.~P.}\
  \bibnamefont {Garrahan}}, \bibinfo {author} {\bibfnamefont {X.}~\bibnamefont
  {Lin}}, \bibinfo {author} {\bibfnamefont {M.}~\bibnamefont {Schr{\"o}der}},
  \bibinfo {author} {\bibfnamefont {N.~R.}\ \bibnamefont {Champness}}, \ and\
  \bibinfo {author} {\bibfnamefont {P.~H.}\ \bibnamefont {Beton}},\ }\href
  {\doibase 10.1126/science.1163338} {\bibfield  {journal} {\bibinfo  {journal}
  {Science}\ }\textbf {\bibinfo {volume} {322}},\ \bibinfo {pages} {1077}
  (\bibinfo {year} {2008})},\ \Eprint
  {http://arxiv.org/abs/http://science.sciencemag.org/content/322/5904/1077.full.pdf}
  {http://science.sciencemag.org/content/322/5904/1077.full.pdf} \BibitemShut
  {NoStop}%
\bibitem [{\citenamefont {Moessner}\ and\ \citenamefont
  {Sondhi}(2001)}]{Moessner2001}%
  \BibitemOpen
  \bibfield  {author} {\bibinfo {author} {\bibfnamefont {R.}~\bibnamefont
  {Moessner}}\ and\ \bibinfo {author} {\bibfnamefont {S.~L.}\ \bibnamefont
  {Sondhi}},\ }\href {\doibase 10.1103/PhysRevLett.86.1881} {\bibfield
  {journal} {\bibinfo  {journal} {Phys. Rev. Lett.}\ }\textbf {\bibinfo
  {volume} {86}},\ \bibinfo {pages} {1881} (\bibinfo {year}
  {2001})}\BibitemShut {NoStop}%
\bibitem [{\citenamefont {Vaezi}\ \emph {et~al.}(2016)\citenamefont {Vaezi},
  \citenamefont {Ortiz},\ and\ \citenamefont {Nussinov}}]{PhysRevB.93.205112}%
  \BibitemOpen
  \bibfield  {author} {\bibinfo {author} {\bibfnamefont {M.-S.}\ \bibnamefont
  {Vaezi}}, \bibinfo {author} {\bibfnamefont {G.}~\bibnamefont {Ortiz}}, \ and\
  \bibinfo {author} {\bibfnamefont {Z.}~\bibnamefont {Nussinov}},\ }\href
  {\doibase 10.1103/PhysRevB.93.205112} {\bibfield  {journal} {\bibinfo
  {journal} {Phys. Rev. B}\ }\textbf {\bibinfo {volume} {93}},\ \bibinfo
  {pages} {205112} (\bibinfo {year} {2016})}\BibitemShut {NoStop}%
\end{thebibliography}%

 \clearpage
 \newpage

 \setcounter{equation}{0}
 \renewcommand{\theequation}{S\arabic{equation}}
 \setcounter{figure}{0}
 \renewcommand{\thefigure}{S\arabic{figure}}

 \appendix
 % \begin{widetext}
 \begin{center}
 \large
 \textbf{
 Supplemental material of ``Classical topological paramagnetism''}
 \normalsize
 \end{center}
 
\section{General definition of a local constraint, confinement, and deconfinement} 
\label{App:Constraint}
Here we address the issue of how one generally defines a lattice
constraint as well as confined and deconfined phases. A local
constraint on a lattice can be abstracted as followed: First one
requires a local mapping from the degrees of freedom to group elements
in $G'$. For the sake of simplicity we take $G'$ abelian. This mapping
should be local such that the value $g_x$ obtained at point $x$
involves degrees of freedom near $x$. Furthermore, it must be neutral
such the product of $g_x$ over a closed manifold yields the
identity. The constraint is then the requirement that $g_x = I$ ($I$
being the identity) at all positions $x$. A defect $f_x$ is a local
violation of this rule in which $g_x = f_x \neq I$. In the familiar
context of $3D$ lattice gauge theories on a cubic lattice, this
mapping would be a mapping between boxes and magnetic charges within
them. A local defect would thus be a particular box where the magnetic
charge is $f$ instead of the identity.

Confined and deconfined phases are defined, as usual, by the free
energy cost $\Delta F_l$ of taking two static opposite defects
($f,f^{-1}$) apart. Confinement is defined by a free energy cost which
increases as a positive power of the distance ($l$) and a deconfined
phase is define by a saturating free energy cost. Just like in the
case of broken symmetries, these define two distinct phases of matter
which can only be connected through a phase transition.  The simplest
way to show this is to remove the constraint and instead introduce
Lagrange multipliers at every point where the constraint is imposed
as:
\begin{align}
  \delta_{f,I} = \frac{1}{|G'|}\sum_{\lambda}\chi_{\lambda}(f)\, ,
\end{align}
where $\lambda$ goes through $|G'|$ values labeling the irreducible
one-dimensional representations of $G'$ and $\chi_\lambda(f)$ is the
character. If $G'=\ZZ_N$, we simply have
$\chi_{\lambda}(f=a^k)=e^{2\pi i \lambda k/N}$, where $a$ is the
generator of $\ZZ_N$.  By the neutrality condition, the resulting
partition function obeys a global symmetry $G'$ shifting all the
$\{\lambda_x\}$ by the same amount. Finally, $\Delta F_l$ is given by
\begin{align}
e^{-\Delta F_l} &= \langle \chi_{\lambda_0}(f)
\chi_{\lambda_{l}}(f^{-1})\rangle\, ,
\end{align}   
and the confined phase translates into the phase with exponentially
decaying $\lambda_x$ correlations (i.e. no spontaneous symmetry
breaking) and the deconfined phase becomes the spontaneous broken
symmetry phase.

\end{document}